\documentclass[acmlarge]{acmart}
\AtBeginDocument{%
  \providecommand\BibTeX{{%
    \normalfont B\kern-0.5em{\scshape i\kern-0.25em b}\kern-0.8em\TeX}}}

\setcopyright{acmcopyright}
\copyrightyear{2026}
\acmYear{2026}
\acmDOI{XXXXXXX.XXXXXXX}
\usepackage[linesnumbered,ruled,vlined]{algorithm2e}
\usepackage{multirow}
\usepackage{xcolor}
\usepackage{listings}

\begin{document}
\title{CFALR: Collaborative Filtering-Augmented Large Language Model for Personalized Fashion Outfit Recommendation}

\author{Yujuan Ding}
\email{dingyujuan385@gmail.com}
\affiliation{
  \institution{The Hong Kong Polytechnic University}
  \country{Hong Kong SAR, China}
}

\author{Junrong Liao}
\affiliation{%
  \institution{University of Electronic Science and Technology of China}
  \country{China}
}

\author{Yunshan Ma}
\affiliation{
  \institution{Singapore Management University}
  \country{Singapore}
}

\author{Yi Bin}
\affiliation{
  \institution{Tongji University}
  \country{China}
}

\author{Wenqi Fan}
\affiliation{
  \institution{The Hong Kong Polytechnic University}
  \country{Hong Kong SAR, China}
}

\author{Tat-Seng Chua}
\affiliation{
  \institution{National University of Singapore}
  \country{Singapore}
}

\author{Qing Li}
\affiliation{
  \institution{The Hong Kong Polytechnic University}
  \country{Hong Kong SAR, China}
}

\begin{abstract}
Personalized outfit recommendation poses a significant challenge in e-commerce and social media platforms, requiring systems that balance user preferences with aesthetic compatibility. Collaborative filtering (CF) provides a traditional solution for this, but it struggles with data-sparse scenarios and complex user-item-outfit relationships. Meanwhile, existing template-based approaches are constrained by rigid pre-designed structures. To bridge these research gaps, we introduce CFALR (Collaborative Filtering-Augmented Large Language Model for Recommendation), a novel framework that synergizes collaborative filtering with large language models for personalized outfit recommendation. Specifically, CFALR describes user-outfit interactions in natural language and leverages LLMs to capture fashion semantics while employing CF-enhanced embeddings to bridge the semantic space and the collaborative interaction spaces. Our technical contributions include: (1) the first LLM-based architecture specifically designed for personalized outfit recommendation, (2) a CF-augmented generative mechanism that efficiently navigates the extensive combination space of outfit items, and (3) trainable projection layers that optimally integrate relational and content features. Experiments on Polyvore and IQON benchmarks demonstrate CFALR's superior performance over both traditional CF-based and LLM-based methods in personalized fill-in-the-blank and personalized outfit generation tasks. 
\end{abstract}


\maketitle

\section{Introduction}
\label{sec:introduction}

Fashion is not only an essential part of people's daily life but also an important way of self-expression~\cite{ding2021leveraging,ma2020knowledge}. Among millions of fashion options, recommender systems can help people to discover items they may like or be suitable to match together, providing personalized services. Therefore, personalized outfit recommendation has become a particularly critical area in e-commerce, as it enhances user satisfaction by tailoring recommendations to individual tastes. Beyond recommending single items, there is growing demand for generating complete and personalized outfits that consider both individual user preferences and aesthetic compatibility of fashion items.

Existing methods for personalized outfit recommendation face several limitations. Most approaches rely heavily on collaborative filtering (CF) to explore interaction patterns from training data~\cite{chen2019pog,song2019gp,ding2023modeling,li2020hierarchical}, while theis type of approaches are sensitive to the quality of user-item interaction data. For example, in scenarios where only sparse or even no interactions (cold-start items) are available, such CF-based approaches will be ineffective or even unusable~\cite{song2019gp}.
Additionally, unlike general personalized recommendation methods that only focus on suggesting individual items, personalized outfit recommendation requires generating complete outfits for different users, involving modeling complex interaction patterns among users, outfits, and items ~\cite{ding2023modeling,yu2018aesthetic}. 
For this problem, previous methods are mostly based on pre-defined outfits~\cite{li2020hierarchical,lu2021personalized,zhan20213}, which cannot keep up with the fast-evolving fashion trends or include new products. 
In comparison, personalized outfit generation methods are able to generate new outfits tailored for different users~\cite{chen2019pog,ding2023personalized}. For example, an existing method~\cite{ding2023personalized} incorporates pre-defined or statistically derived outfit templates as a bridge between personal taste and item compatibility. However, these template-based methods are constrained by their reliance on fixed template sets, which are often incomplete and prone to out-of-distribution issues. 

In parallel, recent advances in Large Language Models (LLMs) have revolutionized natural language understanding and generation~\cite{achiam2023gpt,yang2023dawn}, offering unique capabilities in recommendation in terms of discerning nuanced semantics, exploring diverse user interests, and generalizing to rare or unseen instances~\cite{zhao2023recommender,bao2023tallrec,wang2026mixture,zhao2025webrec}. These strengths position LLMs as promising tools for fashion recommendation. 
By leveraging pre-trained knowledge from massive datasets, LLMs can effectively capture subtle differences in the functional and aesthetic attributes of fashion items. This capability is particularly valuable for addressing cold-start and data-sparse challenges in the fashion domain, where new items emerge much faster than in most other domains~\cite{ding2023computational}.
Furthermore, LLMs offer a unique advantage in tailoring recommendations to individual users, as well as the creation of diverse and creative outfit suggestions.
They can analyze users' functional needs or aesthetic tastes for fashion through the natural language description of user behavior or item content.
However, despite their potential, LLM-based recommendation methods still face inherent limitations. Specifically, although LLMs excel at tasks involving semantic understanding and language generation,  their learned semantic space, which is based on token IDs, is fundamentally misaligned with the user-item interaction spaces required for recommendation tasks, which is typically based on item IDs~\cite{zheng2024adapting}. This misalignment hinders their ability to capture key interaction patterns, often leading to inferior performance compared with traditional CF-based methods. Recent studies have attempted to address this issue by improving the integration of LLMs and CF models, enabling LLM-based recommendation methods to better predict user-item interactions~\cite{kim2024large,wei2024llmrec}. However, these methods are still mainly designed for general personalized recommendation and lack the specific capabilities required for personalized outfit generation, such as outfit-level compatibility modeling and strategic item composition. Therefore, existing LLM-based recommendation methods cannot be directly used as off-the-shelf solutions for our target task.

To overcome these challenges and leverage LLMs' capability and inherent knowledge of fashion understanding for fashion outfit generation, we propose \textbf{C}\textbf{F}-\textbf{A}ugmented \textbf{L}LM for \textbf{R}ecommendation (\textbf{CFALR}). By converting user-outfit interaction behavior into descriptive instances in natural language, we frame the task as a Personalized Fill-In-The-Blank (P-FITB) problem. CFALR bridges the gap between CF-based and LLM-based approaches by employing an open-sourced language model as its backbone recommender and leveraging relational and content features of users and items, which are extracted using foundation visual and CF models. Each non-textual feature is embedded as a single token and embodied in the LLM recommender using a trainable linear projection layer, with separate layers applied for different feature types. We fine-tune the model with the P-FITB objective for tailoring a fashion outfit-specific recommendation model. 
To address another key challenge in outfit generation, namely the extensive search space and combination set~\cite{ding2023personalized}, we propose a CF-augmented generation scheme for inference.
The augmentation is implemented by an output-layer integration, specifically, combining the output probabilities of the CF and language models~\cite{khandelwal2019generalization,fan2024survey}. Experimental results on two benchmark fashion outfit recommendation datasets, \textit{i.e.} Polyvore and IQON, demonstrate the effectiveness of the proposed CFALR method, which outperforms both traditional CF-based and off-the-shelf LLM-based fashion recommendation models on both the personalized fill-in-the-blank (P-FITB) and personalized outfit generation (POG) tasks. 

The main contributions of this paper are as follows:
\begin{itemize}
    \item This paper proposes CFALR, an LLM-based personalized fashion outfit recommendation model. To the best of our knowledge, this is the first time LLMs being applied within this specialized domain of personalized outfit-level modeling. By combining the strengths of LLMs and CF models, the proposed model is able to handle complex user-item-outfit relationships and semantic nuances in the fashion domain.
    \item A CF-augmented generative mechanism is further proposed to facilitate the outfit generation during inference. By synergistically integrating collaborative signals into the LLM-based reasoning results, this design ensures that the generated outfits are not only semantically and stylistically coherent but also strictly grounded in item-level user preferences, effectively bridging the gap between generative flexibility and recommendation accuracy. 
    \item Experimental results on two benchmark fashion outfit datasets demonstrate that the proposed method outperforms existing methods for both the Personalized Fill-In-The-Blank (P-FITB) and Personalized Outfit Generation (POG) tasks. 
\end{itemize}

\section{Related Work}
We review three lines of related work: CF-based recommendation,  LLM-based recommendation methods, and fashion outfit recommendation. 

\subsection{CF-based Recommendation}
Recommender systems have evolved significantly over the years, yet CF-based methods continue to serve as foundational approaches for most application scenarios. 
Traditional CF methods, such as matrix factorization (MF)~\cite{rendle2012bpr}, effectively model user-item interactions by learning latent representations from historical user-item interaction data. With the development of Deep Neural Networks (DNNs), DNN-based CF methods enable more expressive modeling of users and items as well as their interactions~\cite{he2017neural,ding2021leveraging2}. 
However, these methods often struggle with cold-start problems and lack the ability to incorporate contextual information. Later development on hybrid recommendation methods has tried to address these challenges by integrating CF with content features. 
For example, the Factorization Machine (FM)~\cite{rendle2010factorization} method leverages attribute factors of users or items on top of IDs, and models attribute-level interactions. 
Another effective solution to this issue is incorporating multimedia features to enhance item modeling~\cite{he2016vbpr,li2023text}. From another perspective, previous research has applied Graph Neural Networks to improve the exploration of complex interactive relationships. 
For example, NGCF~\cite{wang2019neural} and LightGCN~\cite{he2020lightgcn} model high-order connectivity between users and items based on the bipartite user-item graph. 
In general, traditional CF-based recommendation methods are effective in fitting interaction patterns but are unable to encode detailed semantic information of users or items. Therefore, they are less effective in scenarios where observed interactions are sparse. 
Furthermore, CF methods are inadequate for complex recommendation scenarios, such as those involving combinatorial recommendations or complex conditions, as they lack both the flexibility and generalization ability required to handle such tasks effectively.

\subsection{LLMs for Recommendation}
In recent years, with the development of LLMs, there has been a growing interest in leveraging LLMs for recommender system development~\cite{fan2024survey,zhang2024notellm,liao2024llara}. 
Pioneering studies on LLM-based recommendation relied on In-Context Learning, which directly asks LLMs to make recommendations by using natural language-based prompts~\cite{dai2023uncovering,gao2023chat,qu2026diffusion,sun2023dynamic}. 
However, language models pre-trained on NLP data cannot align well with recommendation tasks~\cite{zhang2025collm}, therefore, more recent methods fine-tune the models on recommendation tasks by transferring user-item interaction data into natural language-described instances. 
These pipelines, however, fail to explicitly model user-item interaction patterns, which are critical for capturing contextual relevance and personalization in recommendation tasks.
To tackle this problem, Zhu et al.~\cite{zhu2024collaborative} proposed CLLM4Rec, which extends the vocabulary of pretrained LLMs with user/item ID tokens to model user/item collaborative and content semantics. With a soft+hard prompting strategy, their model learns the token embeddings via language modeling on RS-specific corpora.

Furthermore, early-stage LLM-based recommendation models heavily rely on semantic priors and tend to overlook collaborative information, making them fall short of traditional methods~\cite{wei2023larger,lin2024rella}. 
To this end, researchers have tried to combine LLMs with collaborative models, for example, integrating collaborative information into the LLM generation process~\cite{bao2023bi,zhang2025collm,zhu2024collaborative,kim2024large}. 
Another line of work tries to use LLMs to offer additional information or knowledge to a standard recommender system. 
For example, LLM-CF~\cite{sun2024large} operates on an offline Chain-of-Thought (CoT) generation paradigm. It utilizes an LLM (specifically RecGen-LLaMA) as an offline component to produce CoT data. 
This data then serves to augment and enhance a separate, traditional Collaborative Filtering (CF)-based recommendation model, primarily for user-item interaction prediction. In essence, the LLM acts as an enhancer for an existing CF framework. KAR~\cite{xi2024towards} transforms open-world knowledge into dense vectors in the recommendation space, compatible with any recommendation models. 

Despite the great success achieved, existing LLM-based methods primarily focus on the general personalized recommendation problem, which typically examines simple second-order user-item interaction, while the research problem in this paper involves complex relationships among users, items, and fashion outfits, therefore cannot be easily addressed by available methods. 
For instance, CLLM4Rec~\cite{zhu2024collaborative} proposes extending the LLM vocabulary with user/item ID tokens for semantic and collaborative understanding, while LLMRec~\cite{wei2024llmrec} utilizes LLMs to enhance standard recommender systems by reinforcing user-item edges, improving item attribute understanding, and aiding user profiling.
In contrast, our problem involves more complex relationships among users, items, and fashion outfits. 
CFALR's objective is to leverage the LLM's inherent prior knowledge of outfit composition alongside CF models' ability to explore hidden interaction patterns from data, integrating multimodal features that depict fashion items. 
We specifically focus on the challenging problem of applying our fine-tuned LLM to generate new outfits for users given very limited input conditions, an area where LLMs typically struggle, as evidenced in our experimental results and discussions in Section~\ref{sec:4.3} and ~\ref{sec:5.8}.

\subsection{Fashion Outfit Recommendation}
Previous research in the area of fashion recommendation has been reviewed systematically and comprehensively~\cite{ding2023computational}. Existing methods for personalized outfit recommendation mostly rely on the idea of collaborative filtering. For example, HFGN~\cite{li2020hierarchical} employed graph neural network to encode the hierarchical structure of user-outfit interactions and outfit-item mappings. POG~\cite{chen2019pog} connects users' preferences regarding individual items and outfits with Transformer architecture. GPBPR~\cite{song2019gp} models item-wise compatibility and personal preference based on the MF-BPR framework. CPTM~\cite{ding2023modeling} integrates the two parts of pattern modeling by applying a single-component translation operation on third-order interactions among the user and item pairs. 
To facilitate efficient personalized outfit generation, Ding et al~\cite{ding2023personalized} proposed incorporating outfit templates into CF models, leveraging users' template preferences as a bridge between personal taste and item compatibility, which also enables to narrow the search space for item combinations. However, it heavily depends on pre-defined or statistically derived template sets dictated by the training data. It also needs to be mentioned that another group of outfit generation methods actually focus on image generation instead of item-wise combination~\cite{yu2025fashiondpo,xu2024diffusion}, which is out of this paper's scope. 

\begin{figure*}[t!]
  \centering
  \includegraphics[width=1\linewidth]{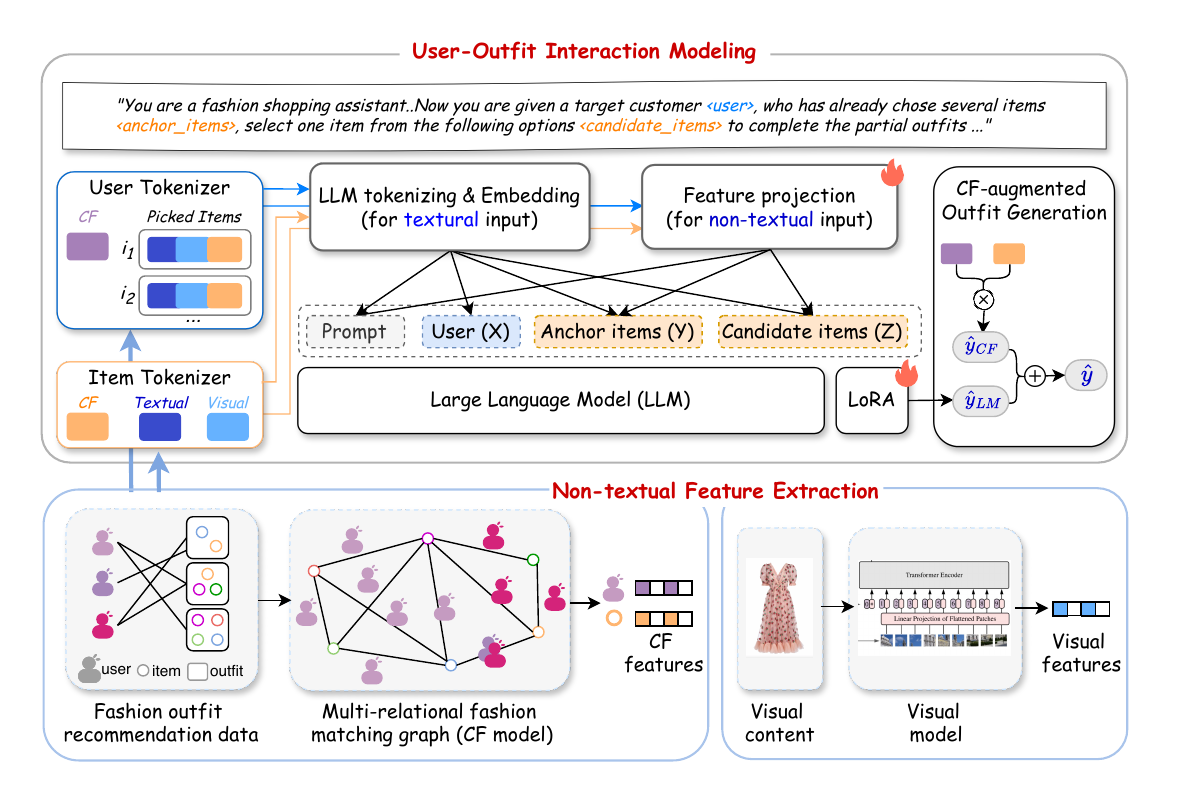}
  \caption{Overview of the CFALR model, which includes four major procedures: 1) non-textual feature extraction, 2) hybrid user and item encoding, 3) user-outfit interaction modeling and 4) LLM recommending. }
  \label{fig:model}
\end{figure*}

Another relevant line of work, bundle generation~\cite{yang2024non,du2023enhancing,nguyen2025ramen,ma2024leveraging,liu2025fine}, aims to provide a bundle of items for the user, which is similar to outfit generation. 
However, existing bundle generation methods are designed for general purpose, therefore, failing to consider the special characteristics of fashion domain. For example, one recent method is proven to be less effective in outfit generation~\cite{2026tudiscrete}, while it's very capable for large-scale bundles such as music list. In addition, most methods focus only on uncovering the combination patterns between items, while ignoring user preference~\cite{hu2023text2bundle}.

To summarize, although fashion recommendation has developed over the years, critical gaps remain. 
Existing outfit generation methods either fail to consider user preferences or rely on pre-defined templates, limiting their ability to deliver personalized and creative outfits. 
From another perspective, integrating the generative capabilities of LLMs with the structured interaction modeling of CF methods has not been fully explored in previous research. 
Our work addresses these gaps by proposing a novel personalized outfit recommendation model that combines CF and LLMs. By introducing a CF-augmented generation mechanism, our approach bridges the strengths of both methods, enabling both recommendation and generation of fashion outfits. 

\section{Problem Formulation}
In this paper we address the personalized fashion outfit recommendation problem, which aims to provide combinative recommendation of fashion items tailored to individual users. Formally, we are given a set of fashion outfits, denoted as $\mathcal{O} = \{o\}$, where each outfit $o=\{i_1, i_2, ... i_{|o|}\}$ consists of multiple fashion items, typically from different categories for functional purposes. Additionally, there is a set of items $\mathcal{I} = \{i\}$ and a set of users $\mathcal{U} = \{u\}$. Each user $u$ interacts with multiple outfits, forming an affiliated outfit set $\mathcal{O}_u$ and an affiliated item set $\mathcal{I}_u$. These interactions are represented by the user-outfit interaction matrix $X_{} = \{x_{u,o} | u \in {\mathcal{U}}, o \in \mathcal{O}\}$. Furthermore, each user-outfit interaction can be further decomposed into interactions between the user and individual items within the outfit, expressed as $x_{u,o} = \{x_{u,i_1}, x_{u, i_2}, ..., x_{u, i_{|o|}}\}$. Each item $i$ is also associated with an image $V_i$, providing rich visual information.

Based on the interaction records between users and outfits, as well as the item combinations within outfits, we aim to develop a personalized fashion outfit recommendation model. Specifically, the problem can be formulated under two common settings: Personalized Fill-In-The-Blank (P-FITB) and Personalized Outfit Generation (POG). 
\begin{itemize}
    \item Personalized Fill-In-The-Blank (P-FITB): In this setting, a target user  $ u \in \mathcal{U}$ and a partial outfit  $\mathcal{I}_p = \{i^p_1, i^p_2, ... \}$, composed of multiple items, are given. The task is to predict the most suitable item  $i_t$ from a candidate set  $\mathcal{I}_c$ to complete the partial outfit while accounting for the personal preferences of $u$. This requires ensuring that $i_t \in \mathcal{I}_u$ and $o_p \cup \{i_t\} \in \mathcal{O}$. 
    
    \item Personalized Outfit Generation (POG): In this scenario, only the target user $u \in \mathcal{U}$ and a single anchor item $i_a \in \mathcal{I}$ are provided. The goal is to select multiple items from the candidate set $\mathcal{I}_c$ to construct a complete outfit for $u$, satisfying the requirements established in the P-FITB task.  
    
\end{itemize}

\section{Method}
In this section, we present the proposed Collaborative Filtering (CF)-Augmented LLM for Recommendation (\textbf{CFALR}). We begin by detailing the core architecture of the model, which is built on large language models (LLMs) and integrates collaborative filtering and content features. Next, we describe the learning strategy employed for training CFALR. Finally, we introduce the CF-augmented generation mechanism utilized during the inference stage for outfit generation.

\subsection{LLM-based fashion recommender architecture}
As introduced in Section~\ref{sec:introduction}, LLMs have been evaluated to have extraordinary capability in context understanding and problem solving at text level. They has also been work that effectively employed as the basic model for different recommendation tasks~\cite{geng2022recommendation, bao2023tallrec, liu2024harnessing}. To solve our problem of personalized outfit generation and recommendation, we follow previous LLM-based recommendation models and apply LLM to build the basic recommender architecture. As illustrated in Figure~\ref{fig:model}, the proposed CFALR model consists of a language model backbone, the hybrid item encoder and the user encoder. 

\subsubsection{LLM as Personalized Outfit Recommender}
We utilize fixed prompt templates to assemble training samples of user-outfit interaction to transform the recommendation problem into a natural language modeling problem which can be processed by an LLM, similar to previous work~\cite{bao2023tallrec,zhang2025collm}. The specific prompt template is illustrated at the left-top part in Figure.~\ref{fig:model}.

\begin{center}
    \noindent\fcolorbox{black}{gray!10}{
    \parbox{.8\linewidth}{\textbf{Prompt Template ($p_0$)}:  You are a fashion shopping assistant whose mission is to help the customers to create their favored fashion outfits. You need to pick one item among the given candidates. Now you are given a target customer <user>, who has already chose several items <anchor\_items>. Please leverage all the information and make a proper choice from the candidate list <candidate\_items> to recommend an item for the customer. The chosen item needs to not only meet the customer’s personal preference, but also match well together (with the given items chose by the customer) to make a fine outfit. }
    }
\end{center}
In the prompt, we describe conditions (input) and the objectives (output) for the specific task. There are three parts of placeholders leaving for the given user, partial outfit and candidate item lists respectively. Specifically, the three placeholders will be filled with hybrid descriptions of corresponding parts respectively: 
$ \pmb{z}_u$ for <user ($u$)>,  
$\{\pmb{z}_{i^p}\}_{i^p \in \mathcal{I}_p}$ for <anchor\_items  ($\mathcal{I}_p$)>, 
and ($\{\pmb{z}_{i^c}\}_{i^c \in \mathcal{I}_c}$ for <candidate\_items ($\mathcal{I}_c$)>. We introduce hybrid encoding below to obtain these hybrid descriptions.

\subsubsection{Hybrid encoding for items and users}
Describing items with natural language and directly applying the textual encoder on it can integrate the item with the language model backbone easily, which is also early approach to designing LLM-based recommendation method. However, there have been studies pointed out that in many cases~\cite{liu2024harnessing}, textual information is not enough to describe the visual details of the item, especially for fashion items with a lot of visual details that cannot be fully expressed by text. Therefore, from a multimedia perspective, our item encoder will incorporate both visual and textual input to fully explore the semantic information of a fashion item. For item $i$, the visual representation can be achieved by $\pmb{v}_i = V(i;\Theta)$, where $V(;\Theta)$ is the visual feature extraction model with $\Theta$ as parameters. However, restricted by the semantic encoding efficiency and always limited input being given, solely semantic encoding is still unable to achieve comprehensive depiction of items. Inspired by related work~\cite{zhang2025collm,zhang2024text}, we propose to leverage multi-relational collaborative filtering encoding to further enhance the modeling of items. Specifically, we adopt a multi-relational CF model, CPTM~\cite{ding2023modeling} to obtain both the item and user representations. We firstly train the CPTM model with triplet data (user-item-item interaction) adapted by the user-outfit interaction data and then adopt the latent embeddings of users and items as the CF features applied in this model. CPTM optimizes the following objective by updating user and item embeddings:
\begin{equation}
    \mathcal{L} = L_{BPR} (s(u, i, j), s(u, i, k)),
\end{equation}
where $L_{BPR}$ denotes the BPR loss and $s(u,i,j)$ denotes the personalized matching score for the user $u$ and a pair of items $i$ and $j$. 
\begin{equation}
    s(u, i, j) = \beta_j - ||\pmb{e}_u + \pmb{e}_i - \pmb{e}_j||^{2} ,
\end{equation}
in which $\pmb{e}_u$, $\pmb{e}_i$ and $\pmb{e}_j$ are  latent embeddings to encode its implicit collaborative features, $\beta_j$ is the bias attached to the target item. 
After training the CPTM model, the latent embeddings of user $\pmb{e}_u$ and item $\pmb{e}_i$ are applied as the CF features of user and item respectively to represent them partially in our CFALR model. 

Overall, as illustrated in Figure~\ref{fig:model}, the hybrid feature list of item $i$ in our method is composed by three parts, \textbf{textual}, \textbf{visual} and \textbf{CF}, ending up as 
\begin{equation}
    x_i = [t_{i,1}, t_{i,2},..., t_{i,K},  \pmb{v}_i, \pmb{e}_i], 
\end{equation}
where $t$ denotes textual token in the textual description and $K$ is the number of tokens of the descriptions. 
Specifically, to encode an item in such a hybrid way in the language model, we apply the following prompt: 
\begin{center}
    \noindent\fcolorbox{black}{gray!10}{
    \parbox{.8\linewidth}{\textbf{Hybrid Item Description ($p^i$)}: <text\_des>. Additionally, we have its interaction history feature <item\_cf\_feat> and its vision feature <item\_vis\_feat>}
    }
\end{center}
While the textual descriptions go through the embedding layer of the LLMs $l()$ and obtain token embedding list $[\pmb{z}^t_{i, 1}, \pmb{z}^t_{i, 2}, ..., \pmb{z}^t_{i, K}]$ ($\pmb{z} = l(t)$), non-textual features are mapped with a projection layer as follows:
\begin{equation}
    \pmb{z}^v = g^v(\pmb{v}), ~~~~\pmb{z}^{cf} = g^{cf}(\pmb{e}),
\end{equation}
where $g^v \in \mathbb{R}^{d_{v} \times d}$ and $g^{cf}\in \mathbb{R}^{d_{cf} \times d}$ are trainable projection layers applying on visual features and collaborative filtering features respectively. $d_v$ and $d_{cf}$ are the dimension of visual and CF features, while $d$ is the dimension of the token embedding of the LLM.  The final hybrid representation for the item is the concatenation of three parts, which turns into:
\begin{equation}
\label{eq:item_emb}
    \pmb{z}_i = [\pmb{z}^t_{i, 1}, \pmb{z}^t_{i, 2}, ..., \pmb{z}^t_{i, K}, \pmb{z}^v_i, \pmb{z}^{cf}_i].
\end{equation}

For users, similar as items, we apply the latent embedding $\pmb{e}_u$ learned by the CPTM model as the collaborative filtering feature. Furthermore, to help encode user's personal preference, we include user's historical interacted items as additional content information. For user $u$ and his interacted item set $\mathcal{O}_u$, due to the input limit of the language model, we sample a subset of interacted items $\pmb{o}_u = \{i^u_1, i^u_2, ..., i^u_M\}$ from the whole set, where $M$ is the length of the user interacted item set. We specifically apply the following prompt to describe users:

\begin{center}
    \noindent\fcolorbox{black}{gray!10}{
    \parbox{.8\linewidth}{\textbf{Hybrid User Description ($p^u$)}: Given a user who has previously picked the following items <item index list>. Additionally, we have her/his interaction history feature <user\_cf\_feat>}
    }
\end{center}

For each item in $i^u \in \pmb{o}_u$, we can obtain its hybrid embedding $\pmb{z}_{i^u}$ based on Eq.~\ref{eq:item_emb}. The final hybrid embedding of the user is the combination of the CF- and interacted item-based embeddings as:
\begin{equation}
    \pmb{z}_u = [\pmb{z}_{i^u_1}, \pmb{z}_{i^u_2}, \pmb{z}_{i^u_M}, \pmb{z}^{cf}_u].
    \label{eq:user_emb}
\end{equation}

\subsection{LLM prediction and optimization}

To effectively train the LLM-based model to understand user's fashion selection and to acquire the capability to capture personal preference and item matching patterns, we reconstruct the personalized outfit recommendation task into the basic conditional choice delivered by textual description, and ask the language model to output the item id they choose in generative language. Specifically, the P-FITB task is applied, with the target user $u$ and his/her selected partial outfit $\mathcal{I}_p=\{i^p_1, i^p_2, ...\}$, and candidate item list $\mathcal{I}_c=\{i^c_1, i^c_2, ...\}$ being the input. 
Each sample will go through the LLM to generate the prediction results. 
We adopt Parameter Efficient Fine-Tuning (PEFT) for the efficient optimization of our LLM-based model since the cost of fine-tuning the entire large language model is too expensive. Specifically, we apply the LoRA~\cite{hu2021lora} method which introduces a small set of additional trainable parameters to the LLM while keeping original LLM parameters fixed. The prediction process thereby is formulated as:
\begin{equation}
    \hat{y} = h_{\hat{\Theta} + \Theta' + \Lambda } (\pmb{Z}),
    \label{eq:pred_llm}
\end{equation}
where $\hat{\Theta}$ and $\Theta'$ denote the fixed parameters of the pre-trained LLM $h( )$ and trainable LoRA parameters respectively, while $\Lambda$ denotes the trainable parameter of the feature projection layers. $\hat{y}$ denotes the predicted item. 
$\pmb{Z}$ is the final embedding list representing the input training sample, which can be derived as follows:
\begin{equation}
    \pmb{Z} = [l(p), [l(p^u), \pmb{z}_u], \{[l(p^u), \pmb{z}_i], i \in \mathcal{I}_p \cup \mathcal{I}_c\}], 
\end{equation}

Our fine-tuning task is to select the most suitable item from the candidate set by comparing the logits of different candidates predicted by the model. 
Formally, our loss function is defined as a cross-entropy loss: 
\begin{equation}
    L = -\frac{1}{N} \sum_{i=1}^N \sum_{c=1}^C y_{i,c} \cdot \log\left(\text{softmax}(\hat{y}_{i,c})\right),
    \label{eq:loss}
\end{equation}
where $N$ represents the number of training samples in the batch size, $C$ represents the number of items in the candidate set. $y_{i,c}$ represents the ground-truth label. 

We apply a multi-stage training strategy to optimize the model considering that directly tuning all modules, including the non-textual projection modules and LoRA parameters, may cause negative interference in the learning process due to the gap between the input feature spaces. 
\begin{itemize}
    \item First Stage (\textbf{S1}): Finetuning the LoRA module ( optimizing $\Theta'$ only with the objective function $\max_{\Theta'}L(\pmb{E}_t; \Theta, \Theta')$) with text-only segment of the prompt (encoded to $\pmb{E}_t$ by the LLM embedding layer), not including non-textual features or additional parameters.
    \item Second Stage (\textbf{S2}): simultaneously finetuning the LoRA and non-textual feature projection parameters, objective function $\max_{(\Theta', \Lambda)}L(\pmb{E}; \Theta, \Theta', \Lambda$).
\end{itemize}

\subsection{CF-augmented outfit generation}
\label{sec:4.3}
Through our experiments and analysis, we found that large language models are good for their understanding ability on complex input, while when the input is short and simple, they might not be able to apply such ability, leading to undesirable performance. For example, in the context of personalized outfit generation, only the user or a single item would be given as the conditions for the generation. Even though we have managed to embed collaborative filtering features to enhance the context encoding of both users and items, the capability of LLMs in modeling simple interaction patterns is still limited, which might be worse than traditional CF model targeted at learning that patterns only. For this consideration, we further propose a collaborative filtering-augmentation mechanism in this paper to tackle this challenge, to help with the outfit generation process at the inference stage. 
More specifically, inspired by the research of Retrieval Augmented Generation (RAG)~\cite{khandelwal2019generalization, fan2024survey}, we propose an output layer-augmentation method, which interpolates the predicted distribution of the CF model and LLM using a hyper-parameter $\lambda$ to produce the final CFALR distribution:
\begin{equation}
    y* = \lambda \hat{y}_{LM} + (1-\lambda) \hat{y}_{CF}. 
    \label{eq:pred}
\end{equation}
Such output-layer augmentation enables to feasibly leverage the strength of both CF and Language models, which is also user-friendly and computationally efficient.

\begin{algorithm}[t]
  \caption{\textbf{Personalized Outfit Generation}}
  \label{alg:1}
\KwIn{ Target User $u$, Initial Anchor Item $i_a$, Candidate Item Set $\mathcal{I}_c$, Maximum Outfit Length $L_o$\\
\textbf{Output:} Generated Outfit $o=\mathcal{I}_y$ \\ 
\textbf{Procedure:}}
Obtain hybrid representation $\pmb{z}_u$ $\pmb{z}_i$ for the user ($u$) and all involved items ($i \in \mathcal{I}_c \cap \{i_a\}$) with Eqn~\ref{eq:user_emb} and Eqn~\ref{eq:item_emb} respectively. \\
Outfit Length $d=1$. \\
Generated Outfit $\mathcal{I}^d_y = \{ i_a\}$. \\
Generated Outfit Score List $S_0=[ ]$. \\
\While{$d \le L_o$}
{
$d = d+1$; \\
Update Anchor Item Set = $\mathcal{I}^{d-1}_y $ \;
Prepare hybrid representation for three parts shown in Prompt Template ($p_0$) \;
Calculate $y^*_{i}$ for $i \in \mathcal{I}_c$ with Eqn.~\ref{eq:pred} and obtain $Y^*_{\mathcal{I}_c}$ \;
Obtain top-ranked candidate item $i^d = \text{Arg}(\text{Max}(Y^*_{\mathcal{I}_c}))$ \;
Update Generated Outfit $\mathcal{I}^d_y = \mathcal{I}^{d-1}_y \cap \{i^d\}$ \;
Update Candidate Item Set $\mathcal{I}_c$ = $\mathcal{I}_c \backslash i^d $ \;
Generated Outfit Score = $y^*_{i^d}$ \;
Update Generated Outfit Score List $S_o.append(y^*_{i^d})$ \;
}
Selected best outfit length $d^* = \text{Arg}\text{Max}(S_o)$ \;
\textbf{Return} Generated Outfit $o = \mathcal{I}^{d^*}_y$\;

\end{algorithm}  

\section{Experiments}
We conduct extensive experiments on two benchmark fashion outfit recommendation datasets to evaluation the proposed method. Existing fashion outfit recommendation methods and LLMs serve as baselines for a comprehensive comparison. We also conduct ablation studies to substantiate our claims in designing the model. With these experiments, we try to address the following research questions:
\begin{itemize}
    \item \textbf{RQ1}: How does CFALR perform on two outfit recommendation tasks compared to conventional? From which perspective are LLMs good at the outfit recommendation problem while from which perspectives are not?
    \item \textbf{RQ2}: Are all technical modules designed in the LLM-based method and the learning strategy for optimizing the model effective? 
    \item \textbf{RQ3}: How does the proposed CFALR perform for the personalized outfit generation problem and how does the proposed CF augmentation mechanism help with that?
\end{itemize}

\subsection{Dataset}
Experiments are conducted on two datasets to evaluate the effectiveness of the proposed CFALR model, which are Polyvore~\cite{ding2023personalized} and IQON~\cite{ding2023modeling}. Both datasets are composed of samples of user-outfit interactions. For Polyvore dataset, we apply the Polyvore-519 version, which has relevantly more users and denser interactions. For IQON dataset, since the original version is large-scale, which results in unaffordable training cost, we sample a subset from the original dataset by keeping 10\% users and 20\% interacted outfits for each user. To construct samples supporting the training of CFALR with the P-FITB task, given a user-outfit interaction, we sample multiple item sets from the outfit and pair with the user, resulting in multiple samples of interactions of user and item pairs.   
For example, if the original sample contains the outfit with $n$ items, after one item is selected as the target output item, we can construct $\sum_{t\in [1, n-2]}\binom{t}{n-1}$ combinations of input and lead to in total  $n \times \sum_{t\in [1, n-2]}\binom{t}{n-1}$ training samples. The statistics of the two datasets can be found in Table~\ref{tab:data}.

\begin{table}
  \caption{Dataset statistics}
  \label{tab:data}
  \begin{tabular}{cccccccccccc}
    \toprule
    ~ &\#I &\#U &\#O &\#U-O &avg.$l(o)$ &$\frac{\#<u,o>}{\#u}$ & $\frac{\#<u,i>}{\#u}$ &$\frac{\#<u,i>}{\#i}$ &\#train &\#$\text{te}_{P-FITB}$ &\#$\text{te}_{POG}$ \\
    \midrule
    Polyvore &19,968 &519 &36,011 &27,012 &2.458 &52.046 &127.952 &3.006 &160,781 &9,561 &9,561\\
    IQON &34,307 &699 &11,026 & 6,612 &3.703 &9.273 &34.339 &1.136 &198,100 &4,414 &4,414\\

  \bottomrule
\end{tabular}
\end{table}

\subsection{Experimental settings}
For training, we leverage the Vicuna-7B-v1.3 model as our large language model backbone and a pre-trained CP-TransMatch~\cite{ding2023modeling} model as our collaborative filtering model. To stay consistent with previous outfit recommendation methods for the sake of a fair comparison, we apply ResNet-50 as our visual model to extract visual representations for items. The hidden size of collaborative filtering embeddings is 32, and 2048 for visual embeddings. We use AdamW optimizer in both tuning stages, with linear warm-up and cosine annealing strategy~\cite{zhang2025collm}. The learning rate starts at $10^{-5}$ and increases linearly to $10^{-3}$ during the warm-up phase. Afterward, it decreases following a cosine curve to the $8\time 10^{-6}$. Our training batch size is set to 4. For the LoRA module, we set $r$, \textit{alpha} and \textit{dropout} to 8, 16, 0.05 and the target module “[q\underline{\phantom{X}}proj, v\underline{\phantom{X}}proj]”. 
For the Polyvore dataset, both training stages (first and second) run for three epochs. For the IQON dataset, the first-stage training takes three epochs while the second-stage training takes one epoch. All experiments are conducted on a server with 8 NVIDIA RTX A6000 GPUs.

During inference, for P-FITB, each test sample is composed of one user and an outfit with several items. For each sample, we randomly select one item from the outfit as the prediction target and replace it with a blank mask. Furthermore, we randomly sample $k-1$ items that the user is not fond of from the whole item set as the negative items to construct the candidate item set with $k$ items. We set $k\in \{4, 10, 20\}$ so that different levels of selection difficulties can be explored. In comparison, for POG, for each user-outfit interaction record, we select one item from the outfit as the given condition and leave the rest items as targets for generation. Meanwhile, we pair each test sample with multiple negative items and make the candidate item set with the length of $20$. The maximum outfit length is set to 5. Number of historical interacted items are set to 5 for Polyvore and 3 for IQON to optimize balance between accuracy and computational efficiency. 

\subsection{Evaluation}
The evaluation of the proposed model includes two tasks: Personalized Fill-in-the-blank (P-FITB) and Personalized Outfit Generation (POG). 
\subsubsection{Personalized Fill-in-the-blank}
We employ \textbf{Accuracy (\textit{Acc})} to evaluate the performance of various models for the P-FITB task, which measures the percentage of the correct prediction over all predictions as follows:
\begin{equation}
    Acc = \frac{N_r}{N},
\end{equation}
where $N_r$ and $N$ denotes the number of correctly predicted and all testing samples respectively. 

\subsubsection{Personalized Outfit Generation}
Evaluating Personalized Outfit Generation (POG) presents significant challenges compared to traditional discriminative tasks, primarily due to the open-ended and subjective nature of generative fashion modeling. To rigorously assess the performance of CFALR and baseline models, we establish a multi-dimensional evaluation framework consisting of three complementary methods: (1) Quantitative Overlap Analysis, which measures the alignment with ground-truth outfits to assess recommendation fidelity; (2) LLM-as-a-Judge, leveraging the semantic reasoning of large models to evaluate stylistic quality of outfits; and (3) Human Evaluation, providing a human assessment on the generated results. The specific configurations of these evaluations are detailed below.

Given the user $u$ and the item $i$, the generated outfits $o_{ui}$ have two-perspective requirements, which corresponds to the two objectives of the generation task. First, the generated outfit needs to meet the personal preference of the given user (\textbf{Personalization}), which can be measured by counting the items in the outfit favored by the user, i.e., positive items to the user. We define the metric of \textbf{P}ersonalization \textbf{P}recision (\textbf{\textit{PP}}) to measure this perspective of performance, which can be calculated as follows:

\begin{equation}
    PP(o_{ui}) = \frac{\left | o_{ui} \cap  \mathcal{I}_u  \right | }{\left |o_{ui}  \right | }, 
\end{equation}
where $\mathcal{I}_u$ is the positive item set to $u$, consisting items that the user likes. 

Second, items in the generated outfits need to match well with each other and all together (\textbf{Compatibility}). We measure this perspective of performance by comparing how much the generated outfit overlaps with the pre-defined ones, which are assumed to be well-matched. This can be further implemented in two levels, outfit-level and pairwise-level. The \textbf{O}utfit \textbf{C}ompatibility (\textbf{\textit{OC}}) is defined as follows:
\begin{equation}
    OC(o) = \max_{\hat{o} \in \mathcal{O}}{J(o, \hat{o})},
\end{equation}
where $o$ is the generated outfit and $\hat{o}$ is the pre-defined outfit. 
$J(\cdot)$ is the Jacobian Similarity:
\begin{align}
J(A,B)=\frac{\left | A\bigcap B  \right | }{\left | A\bigcup B \right | }, 
\end{align}
which is usually used to measure the similarity between two sets. 
The pairwise-level compatibility, i.e., \textbf{M}ean \textbf{P}airwise \textbf{C}ompatibility (\textit{\textbf{MPC}}) is defined as the average of pairwise compatibility among an outfit, which based on the matching measurement of two items:
\begin{equation}
    m(i,j) = 1~~ if (i,j) \in \mathcal{P}~~ else~~ 0, 
\end{equation}
$\mathcal{P}=\{(k,l)\}$  is the positive item pair set in which $(k,l)\in o$ for $o \in \mathcal{O}$.
Then the MPC is defined as:
\begin{equation}
    MPC(o) = \frac{\sum_{(i,j) \in \mathcal{P}_o}{m(i,j)}}{|\mathcal{P}_o|}, 
\end{equation}
where $\mathcal{P}_o$ is the pair set of the generated outfit $o$. $\mathcal{P}_o=\{(i,j) \in o\} \cap \{i \neq j\}$ is the item pair set of the outfit $o$.

The LLM- and human-based evaluations are designed to complement the quantitative overlap analysis, particularly in assessing the quality of novel generated outfits that may not exist in the `ground truth'. Specifically, we task both LLMs and human experts with evaluating a sampled subset of generated outfits across three key dimensions: \textit{completeness}, \textit{compatibility}, and \textit{aesthetics}. For each criterion, a numerical score is assigned within the range of $[0, 5]$ to provide a fine-grained qualitative assessment. In the end, we calculate the average comparable score for each method to obtain its overall performance in terms of generated outfit quality.

\subsection{Baselines}
Our focus is to compare CFALR with models specifically tailored or readily adaptable to the complexities of outfit generation. This approach ensures a more relevant and accurate evaluation of CFALR's unique contributions to this specific domain. Our experiments included various types of methods to deliver extensive comparison and analysis, including CF-based recommendation methods, off-the-shelf LLM-based methods, advanced Vision-Language Models (VLMs) and bundle recommendation methods. All compared methods are introduced below:
\begin{itemize}
    \item \textbf{GP-BPR}~\cite{song2019gp} is a personalized item matching method that decomposes the problem into the modeling of personal preference and general item compatibility. These two parts are modeled by MF independently.  
    
    \item \textbf{DGSR}~\cite{ding2021leveraging} is a method originally for sequential fashion recommendation but also for the modeling of third-order interaction data between user and item pair. It is built upon two graphs to enhance the user-item interaction and item-item transition modeling. 
    
    \item \textbf{CPTM} (CP-TransMatch)~\cite{ding2023modeling} is a recent personalized item matching method which applies a single-component translation operation to model the user-item pair interaction.  
    
    \item \textbf{Vicuna}~\cite{chiang2023vicuna} is an open-sourced LLM  fine-tuning LLaMA on user-shared conversations collected from ShareGPT, which is also the backbone of our method. 
    
    \item \textbf{Qwen-VL Series} (\textbf{Qwen2.5-VL}~\cite{bai2025qwen2}, \textbf{Qwen3-VL}~\cite{yang2025qwen3}, and \textbf{Qwen3.5-VL}~\cite{qwen3.5}) are state-of-the-art open-sourced MLLMs designed for integrated vision-language understanding. Compared to text-centric LLMs, these models are superior for multimodal tasks as they allow for the direct encoding of item images without the intermediate step of textual attribute extraction. We incorporate three iterations of the Qwen-VL series as baselines to assess the impact of varying multimodal model scales and architectures on our proposed tasks.
    \item \textbf{GPT Series} (\textbf{GPT-4}~\cite{achiam2023gpt}, \textbf{GPT-5.4-mini}~\cite{OpenAIGPT5_4}) are representative advanced closed-sourced LLMs developed by OpenAI which excels in many tasks. We apply two versions of GPT models, including the text-only GPT-4 and the MLLM version GPT-5.4.
    
    \item \textbf{LLMCBR}~\cite{liu2025llmcbr}  is a bundle recommendation framework that leverages large language models to get bundle-level semantic representations. It integrates multi-view and multi-grained collaborative information by modeling user-item and user-bundle interactions across different views, enabling the capture of both global and local user preferences.
    \item \textbf{Bundle-MLLM}~\cite{liu2025fine} is a multimodal large language model framework for product bundling that incorporates textual, visual, and relational features through a hybrid feature fusion strategy. It reformulates bundling as a multiple-choice question and adopts a progressive optimization scheme to enhance both bundle pattern learning and multimodal semantic understanding.
    \item \textbf{TOG}~\cite{ding2023personalized} is a personalized outfit generation method based on pre-defined templates. We only apply this method for POG task in the experiments. 

\end{itemize}

It should be noted that our primary objective with CFALR is personalized fashion outfit generation, a distinct and more complex task compared to the general personalized item recommendation problems. Applying item recommendation methods to our task would necessitate substantial adaptation, requiring entirely new modules for outfit-level compatibility modeling and item composition strategies. Such extensive modifications would obscure their true performance, as results would depend heavily on our custom extensions rather than their original design. Therefore, existing LLM-based item recommendation methods were not included as direct baselines, as our focus remained on models specifically tailored or readily adaptable to the complexities of outfit generation, ensuring a relevant and accurate evaluation of CFALR's unique contributions to this specific domain.

\begin{table}
  \caption{Personalized Fill-In-The-Blank (P-FITB) Performance (Acc)}
  \label{tab:fitb_overall}
  \begin{tabular}{c|c|ccc|ccc}
    \toprule
    \multicolumn{2}{c|}{Dataset}  &~ &Polyvore &~  &~ &IQON  &~ \\
    \cmidrule(r){1-8}
    Method Type  &Method &1/4 &1/10 &1/20 &1/4 &1/10 &1/20\\
    \cmidrule(r){1-8}
    
    \multirow{3}{*}{CF-based Rec} &GP-BPR~\cite{song2019gp} &0.4107	&0.2599	&0.1923&0.3626 &0.1772 &0.1060 \\
    &DGSR~\cite{ding2021leveraging} &0.3538	&0.2034	&0.1621 &0.2626 &0.1232 &0.0736 \\
    &CPTM~\cite{ding2023modeling} &0.4320	&\underline{0.2744}	&\underline{0.2064}  &0.3002 &0.1484 &0.0940 \\
    \cmidrule(r){1-8}
    
    \multirow{2}{*}{LLM} &Vicuna~\cite{chiang2023vicuna} &0.2416	&0.0973	&0.0513 &0.2451 &0.0961 &0.0496\\
    &GPT-4~\cite{achiam2023gpt} &0.3873 &0.1800 &0.0623 &0.3795 &0.2057 &\underline{0.1170}\\
    \cmidrule(r){1-8}
    
    \multirow{4}{*}{VLM} &Qwen2.5-VL-7B~\cite{bai2025qwen2} &0.2903 &0.1260 &0.0687 &0.2693 &0.0938 &0.0679\\
   & Qwen3-VL-8B~\cite{yang2025qwen3} 	& 0.3458	& 0.1331	& 0.0732	& 0.3309	& 0.1460	& 0.0623 \\
   & Qwen3.5-VL-35B~\cite{qwen3.5} & 0.4746&	 0.2041&	 	0.0913&	 	0.5653&	 	0.2661&	 0.1194 \\
   & GPT-5.4-mini~\cite{OpenAIGPT5_4}	& 0.4426	& 0.1844 	& 0.0672	& 0.48258	& 0.2257	& 0.0711 \\
   \cmidrule(r){1-8}
   
    \multirow{2}{*}{Bundle Rec} & LLMCBR~\cite{liu2025llmcbr} & 0.4371	& 0.2430	& 0.1611	& 0.2825	& 0.1437	& 0.0920 \\
   & Bundle-MLLM~\cite{liu2025fine} & \underline{0.4775}	& 0.2519	& 0.1459 & \underline{0.5214}	& \underline{0.2336}	& 0.1164\\
   \cmidrule(r){1-8}
   
    Ours &CFALR &\textbf{0.6498}	&\textbf{0.3957}  &\textbf{0.2459} &\textbf{0.6103} &\textbf{0.3654} &\textbf{0.2018}\\
    
  \bottomrule
\end{tabular}
\end{table}

\subsection{Overall Performance }
\subsubsection{Personalized Fill-In-The-Blank (P-FITB)}
We present the experimental results on P-FITB of all compared methods in Table~\ref{tab:fitb_overall}, from which we can observe the following results:
\begin{itemize}

    \item Among traditional recommendation baselines, CF-based methods still achieve competitive performance, especially on Polyvore. In particular, CPTM performs strongly under the more challenging 1/10 and 1/20 settings on Polyvore, achieving 0.2744 and 0.2064, respectively. This indicates that collaborative filtering and task-specific recommendation signals remain useful for personalized fill-in-the-blank outfit recommendation. However, their performance is less stable across datasets, as their results on IQON are generally weaker than those of more advanced multimodal models. This may be because Polyvore contains denser user-item interaction records. Such denser interaction data can better support CF-based models in discovering reliable interactive patterns from historical behaviors, whereas sparser interactions in IQON may limit their effectiveness.

    \item When comparing the performances of Vicuna and GPT-4, it is evident that GPT-4 achieves significantly better results overall. Vicuna, on the other hand, shows the weakest performance among all included methods, which underscores the substantial difference in capability between these two large language models (LLMs). GPT-4's strong performance demonstrates its ability to handle domain-specific and customized tasks effectively. However, our CFALR model, built upon Vicuna, surpasses GPT-4 for the targeted task due to the integration of effective technical modules and a robust learning process. This showcases CFALR’s ability to significantly enhance the base model’s capacity to address specialized challenges. 

    \item Advanced VLMs show clear progress compared with earlier LLM baselines. Vicuna performs relatively poorly across all settings, while GPT-4 achieves much stronger results, especially on IQON. More recent VLMs further improve the performance. In particular, Qwen3.5-VL demonstrates strong performance on both datasets and achieves the best baseline results on IQON across all three settings. GPT-5.4-mini also performs competitively, especially under the 1/4 setting. These results suggest that recent advances in multimodal foundation models can substantially benefit personalized outfit recommendation, as stronger visual-language understanding helps the models better capture item compatibility and user-oriented matching signals.

    \item Bundle recommendation models, i.e., LLMCBR and Bundle-MLLM, also show strong competitiveness. Bundle-MLLM achieves the best baseline result on Polyvore under the 1/4 setting and performs strongly on IQON as well. LLMCBR also delivers competitive results, particularly on Polyvore. Although these methods are based on LLM backbones that are generally less advanced than the latest VLMs in terms of general multimodal capability, their strong performance indicates the importance of task-specific learning for bundle and outfit recommendation. This suggests that curated learning on recommendation-oriented data can effectively enhance model performance on high-level and complex personalized outfit matching tasks, beyond relying only on general multimodal reasoning ability.

    \item The proposed CFALR consistently achieves the best performance across all datasets and evaluation settings. Compared with the strongest baseline in each setting, CFALR achieves clear improvements, especially under the more challenging 1/10 and 1/20 settings. These results demonstrate that CFALR can more effectively model personalized preference and outfit compatibility, confirming its robustness and superiority for personalized fill-in-the-blank outfit recommendation.
    
\end{itemize}

\begin{table}
  \caption{Personalized Outfit Generation Performance}
  \label{tab:pog}
  \begin{tabular}{c|c|ccc|ccc}
    \toprule
    \multicolumn{2}{c|}{Dataset}  &~ &Polyvore &~  &~ &IQON  &~ \\
    \cmidrule(r){1-8}
    Method Type  &Method &PP &OC &MPC &PP &OC &MPC\\
    \cmidrule(r){1-8}
    \multirow{3}{*}{CF-based Rec} &GP-BPR~\cite{song2019gp} &0.1649	&0.3942 &0.2137 &\underline{0.2608} &0.3004 &0.2188 \\
    &DGSR~\cite{ding2021leveraging} &0.1645	&0.4193	&0.2457 &0.1906 &0.2790 &0.1801 \\
    &CPTM~\cite{ding2023modeling} &0.2314	&\underline{0.4506}	&\underline{0.3365}  &0.2427 &0.2934 &\underline{0.2234} \\
    \cmidrule(r){1-8}
    \multirow{2}{*}{LLM} &Vicuna~\cite{chiang2023vicuna} &0.0488 &0.3621 &0.0856 &0.1389 &0.2662 &0.1333 \\
    &GPT-4~\cite{achiam2023gpt} &0.0847 &0.2951 &0.0794 &0.1946 &0.2754 &0.1204\\
    \cmidrule(r){1-8}
    \multirow{4}{*}{VLM} &Qwen2.5-VL-7B~\cite{bai2025qwen2} &0.0944 &0.3800 &0.1264 &0.1411 &0.2660 &0.1370\\
    & Qwen3-VL-8B~\cite{yang2025qwen3} & 0.1317	& 0.2750	& 0.0663	& 0.1917	& 0.2594	& 0.0915\\
    & Qwen3.5-VL-35B~\cite{qwen3.5} & 0.2106 & 0.4106 & 0.2075 & 0.2524	& \underline{0.3036}	& 0.1407\\
    & GPT-5.4-mini~\cite{OpenAIGPT5_4} & 0.2202	& 0.4218	& 0.2192	& 0.1084	& 0.2598	& 0.1085\\
    \cmidrule(r){1-8}
    \multirow{2}{*}{Bundle Rec} 
    & LLMCBR~\cite{liu2025llmcbr} & 0.2281	& 0.4331	& 0.3151	& 0.2317	& 0.2774	& 0.2097\\
    & Bundle-MLLM~\cite{liu2025fine} & \underline{0.2337}	& 0.4361	& 0.2869 & 0.2578	& 0.2966	& 0.1933\\
    \cmidrule(r){1-8}
    Outfit Gen &TOG~\cite{ding2023personalized}  &0.1399	&0.3894	&0.1986 &0.1832 &0.2612 &0.1364 \\
    \cmidrule(r){1-8}
    Ours &CFALR &\textbf{0.2428}	&\textbf{0.4626}	&\textbf{0.3556} &\textbf{0.2883} &\textbf{0.3147} &\textbf{0.2291}\\

  \bottomrule
\end{tabular}
\end{table}

\subsubsection{Personalized Outfit Generation (POG)}
Furthermore, we present the performance of the proposed CFALR and all compared methods on the Personalized Outfit Generation (POG) task in Table~\ref{tab:pog}. This part of experimental results further highlight the effectiveness and versatility of the proposed method in personalized fashion outfit recommendation, which showcase the strengths of proposed CFALR in personalization and compatibility modeling by comparing with both CF-based and LLM-based methods. We specifically have the following observations and analysis. 
\begin{itemize}
    
    \item Overall, CFALR achieves the best performance across all datasets and evaluation metrics. These results show that CFALR can effectively capture user preferences while maintaining strong item compatibility, which is crucial for generating high-quality personalized outfits. Compared with the strongest baseline in each setting, CFALR still achieves clear improvements, demonstrating its robustness across both datasets.

    \item CF-based recommendation methods remain highly competitive on the POG task. In particular, CPTM performs strongly on Polyvore, achieving 0.2314, 0.4506, and 0.3365 on PP, OC, and MPC, respectively, and also obtains the strongest baseline result on IQON in terms of MPC. GP-BPR also performs well on IQON, especially on PP. This indicates that collaborative filtering methods can effectively exploit historical user-item and item-item interaction patterns for outfit generation. Compared with the P-FITB task, the performance gap between CFALR and CF-based methods is less pronounced in POG. This is because POG provides relatively limited input and focuses more on directly generating a complete outfit composition, where interaction-pattern modeling plays an important role.

    \item Pure LLM-based methods, including Vicuna and GPT-4, generally underperform compared with CF-based and specialized recommendation methods. Although LLMs have strong language understanding and reasoning capabilities, the POG task requires direct item selection and outfit composition rather than understanding complex textual or visual input. As a result, pure LLMs are less effective in modeling personalized outfit compatibility and user-item interaction patterns. This observation is consistent with the results of the P-FITB task: LLM-based methods are more useful when rich contextual input needs to be interpreted, while their advantages become less evident when the task mainly requires structured data modeling for recommendation and generation.

    \item General-purpose VLMs show stronger performance than pure LLMs, but their results are still inconsistent across datasets and metrics. For example, Qwen3.5-VL achieves competitive performance on both datasets, especially on IQON, where it obtains the strongest OC result among all baselines. GPT-5.4-mini also performs well on Polyvore, achieving strong PP and OC scores. These results suggest that advanced multimodal foundation models can benefit fashion recommendation by leveraging visual-language understanding. However, their performance does not consistently surpass task-specific recommendation models. For instance, GPT-5.4-mini performs strongly on Polyvore but is less competitive on IQON, and Qwen3-VL does not always improve over Qwen2.5-VL. This indicates that general multimodal reasoning ability alone is insufficient for fully solving personalized outfit generation, where recommendation-specific learning and compatibility modeling remain essential.

    \item Bundle recommendation methods, i.e., LLMCBR and Bundle-MLLM, achieve strong and stable performance across the two datasets. Bundle-MLLM obtains the best baseline PP score on Polyvore and also performs competitively on IQON, while LLMCBR achieves strong MPC performance on Polyvore. Their strong results show that bundle recommendation is highly relevant to personalized outfit generation, as both tasks require modeling a set of items as a coherent whole. Although these methods may not rely on the most advanced general-purpose large model backbones, their task-specific learning enables them to better capture bundle-level compatibility and personalized preference patterns. This further suggests that curated recommendation training is important for high-level outfit generation tasks.

    \item The outfit generation baseline TOG achieves reasonable but relatively limited performance compared with CF-based and bundle recommendation methods. This suggests that general outfit generation models can capture certain compatibility signals, but may not sufficiently model personalized preference. Since POG requires both outfit compatibility and user-specific personalization, methods that only focus on general outfit construction are less competitive than models designed to jointly consider user preference and item compatibility.

    \item Comparing the two datasets, we observe that CF-based methods are particularly strong on Polyvore, while advanced VLMs and bundle recommendation methods become more competitive on IQON. This may be related to the different characteristics of the two datasets. As shown in Table~\ref{tab:data}, IQON outfits contain more items on average than Polyvore outfits, meaning that models need to select more items to construct a complete outfit. In our experiments, LLM- and VLM-based methods tend to select more items when composing outfits, which may partially explain their relatively stronger competitiveness on IQON. Nevertheless, CFALR consistently outperforms all baselines on both datasets, showing that it can adapt well to different outfit composition scenarios.

\end{itemize}

We further report LLM-as-a-judge and human evaluation results on a subset of generated outfits (70 samples), assessing overall quality, as shown in Table~\ref{tab:outfit_eval}. For the LLM-as-a-judge evaluation, we employ two advanced LLMs, namely Gemini-3 and Qwen-3.6. For human evaluation, we invite three fashion experts to manually score each case for each compared method, and report the averaged results in the table. The results show that CFALR consistently outperforms the two baselines, i.e., CPTM and fine-tuned Vicuna, across all evaluation settings. Its superiority is particularly pronounced in terms of compatibility and aesthetics.

\begin{table}
  \caption{Generated Outfit Quality Evaluation by LLMs and Human Expert.}
  \label{tab:outfit_eval}
  \begin{tabular}{c|c|cccc}
    \toprule
     Evaluation Model&Outfit Generation Model &Completeness	&Compatibility	&Aesthetics &Overall\\
      \cmidrule(r){1-6}
    \multirow{3}{*}{Gemini-3} &CPTM &0.531 &0.591 	&0.491 	&0.537 \\
     &Vicuna-T &0.744 	&0.461 	&0.399 	&0.535 \\
     &CFALR &\textbf{0.832} 	&\textbf{0.883} 	&\textbf{0.811} 	&\textbf{0.842}\\
    \cmidrule(r){1-6}
    
    \multirow{3}{*}{Qwen-3.6} &CPTM &0.560	&0.493	&0.483 &0.511\\
     &Vicuna-T &0.597	&0.635	&0.629 &0.621\\
     &CFALR &\textbf{0.645}	&\textbf{0.696}	&\textbf{0.680} &\textbf{0.674}\\
    \cmidrule(r){1-6}

    \multirow{3}{*}{Human Expert} &CPTM &0.563	&0.501	&0.491	&0.518\\
     &Vicuna-T &0.747	&0.709	&0.688	&0.723\\
     &CFALR &\textbf{0.755}	&\textbf{0.747}	&\textbf{0.795}	&\textbf{0.765}\\
     
  \bottomrule
  
\end{tabular}
\end{table}

\begin{table}
  \caption{Ablation study results on CFALR for P-FITB task. \textit{V} denotes visual features, \textit{CF} denotes collaborative filtering features, \textit{UH} denotes user's historical interacted items}
  \label{tab:feat}
  \begin{tabular}{c|ccc|ccc}
    
    \toprule
    \multirow{2}{*}{Ablation Models} &~ &Polyvore &~  &~ &IQON  &~ \\
     \cmidrule(r){2-7}
     &1/4 &1/10 &1/20 &1/4 &1/10 &1/20 \\
    \cmidrule(r){1-7}

    CFALR w/o V\&CF &0.5820	&0.3198	&0.1851 &0.6051 &0.3496 &0.2021\\ 
    CFALR w/o CF &0.5896 &0.3279 &0.1873 &0.6160 &0.3692 &0.2121\\
    CFALR w/o V &0.6142	&0.3788	&0.2502 &0.6123 &0.3679 &0.2091\\
    CFALR w/o UH &0.6264 &0.3701 &0.2207 &0.6012 &0.3581 &0.1991\\
    CFALR &\textbf{0.6498}	&\textbf{0.3957}  &\textbf{0.2459} &\textbf{0.6103} &\textbf{0.3654} &\textbf{0.2018}\\
  \bottomrule
  
\end{tabular}
\end{table}

\subsection{On hybrid representation}
In this part, we investigate the effectiveness of the hybrid encoding from both the item and user sides with the P-FITB task on two evaluation datasets. The results of different settings are reported in Table~\ref{tab:feat}, which demonstrate the following key observations:
\begin{itemize}
    \item \textbf{Effectiveness of Hybrid Representations}: Experimental results across both datasets demonstrate the effectiveness of incorporating three components of features—historical interaction, visual, and collaborative filtering (CF). Including all these features consistently improves performance, validating the design of the hybrid representation.

    \item \textbf{Polyvore Dataset Discussion}: On the Polyvore dataset, CF features play a critical role in enhancing personalized recommendation performance, particularly for more challenging cases like 1/20 selection. Removing CF features results in a substantial performance drop (from 0.2459 to 0.1873, a 28.6\% decrease). Visual features of items and historical interaction content for users also contribute, but their effectiveness varies by task difficulty. User’s historical interactions are more impactful for difficult tasks (e.g., 1/20 selection), while item visual features are more relevant for easier tasks (e.g., 1/4 selection). These findings underscore the importance of complementing LLMs with CF information to enhance recommendation performance. Leveraging CF features derived from well-trained CF models is more effective than relying solely on historical interaction records, which would require the language model to infer personal preferences without guidance. This hybrid design not only aids the language model in understanding personal preferences but also provides clues on item compatibility patterns.

    \item \textbf{IQON Dataset Discussion}: The hybrid representation’s effectiveness is less pronounced on the IQON dataset compared to Polyvore. Among the additional features, user’s historical interactions have the most impact, whereas visual and CF features show limited effectiveness. This outcome is unsurprising given the extreme sparsity of IQON’s interaction data. Referring to Table 1, each item is interacted with an average of only 1.136 times, indicating that most items appear only once in the training data. Such sparsity hampers traditional CF-based methods from capturing interaction patterns, rendering CF features less effective. However, with LLM as the backbone, the inclusion of user interaction data enhances the model’s ability to capture personal preferences, improving overall performance. These results demonstrate the feasibility of LLM-based recommendation methods in scenarios with extreme data sparsity, where traditional CF-based approaches struggle.
\end{itemize}

\begin{table}
  \caption{P-FITB performance of models with various training objectives}
  \label{tab:loss}
  \begin{tabular}{c|cccc|cccc}
    \toprule
    \multirow{2}{*}{Loss Design} &\multicolumn{4}{c|}{Polyvore} &\multicolumn{4}{c}{IQON} \\
    \cmidrule(r){2-9}
    &1/4 &1/10 &1/15  &1/20 &1/4 &1/10 &1/15 &1/20 \\
    \cmidrule(r){1-9}
    
    CFALR loss &\textbf{0.6329}	&\textbf{0.3893}  &\textbf{0.3117} &\textbf{0.2411} &\textbf{0.6103} &\textbf{0.3654} &\textbf{0.2680} &\textbf{0.2018}\\
    LM loss &0.6247 &0.3803 &0.2920 &0.2361 &0.5779 &0.3387 &0.2447 &0.1846\\

  \bottomrule
\end{tabular}
\end{table}

\begin{table}
  \caption{P-FITB performance of models with various training strategies}
  \label{tab:training}
  \begin{tabular}{c|cccc|cccc}
    \toprule
    \multirow{2}{*}{Training Strategy} &    \multicolumn{4}{c|}{Polyvore} &\multicolumn{4}{c}{IQON} \\
    \cmidrule(r){2-9}
     &1/4 &1/10 &1/15  &1/20 &1/4 &1/10 &1/15 &1/20 \\
    \cmidrule(r){1-9}
    \textit{T\_1} &0.5934	&0.3505	&0.2738 &0.2191 &0.6039 &0.3384 &0.2573 &0.1980\\ 
    \textit{T\_2}  &\textbf{0.6498}	&\textbf{0.3957}	&0.3088 &\textbf{0.2459} &\textbf{0.6103} &\textbf{0.3654} &\textbf{0.2680} &0.2018\\
    \textit{T\_3} &0.6329	&0.3893	&\textbf{ 0.3117} &0.2411 &0.5990 &0.3425 &0.2532 &\textbf{0.2021}\\
    \textit{T\_4} &0.4811	&0.0967	&0.0606 &0.0466 &0.5610 &0.2961 &0.1927 &0.1513\\
    \textit{T\_5} &0.6263	&0.3683	&0.2871 &0.2223 &0.5968 &0.3449 &0.2424 &0.1910
    \\
  \bottomrule
\end{tabular}
\end{table}

\subsection{On learning objective and strategy}
In this part, we investigate the impact of different learning objectives and training strategies on the final performance of the proposed model. Specifically, we tested our model with two types of learning objectives, 1) CFALR loss: discriminative loss in Eq.~\ref{eq:loss} and 2) LM loss $L = -\sum_{t=1}^{|y|} \text{log}P(y_t|y_{<t}, x)$ and $P(y_t)$ denotes the probability distribution of the generative token over the whole corpus. The results, presented in Table~\ref{tab:loss} show that CFALR outperforms LM loss across all selection settings and datasets, particularly excelling in higher-difficulty scenarios (e.g., 1/20 in IQON: 0.2018 vs. 0.1846, a 9.3\% improvement). This suggests CFALR's ability to focus on localized decision-making leads to superior accuracy and robustness. This makes it a more effective objective function for the P-FITB task compared to LM loss. Fundamentally, CFALR is a discriminative loss that applies cross-entropy loss on a restricted candidate item set, focusing the model's attention on a smaller, contextually relevant subset of items. This targeted optimization allows the model to better capture fine-grained relationships within the candidate set. From this analysis, we can derive that CFALR is well-suited for traditional recommendation tasks, which is fundamentally discriminative tasks such as ranking or item selection within a predefined candidate set (e.g., personalized product recommendations, where the model must choose from a curated set of items).

To further investigate training strategy, we arrange different training settings and test the evaluation results, reported in Table~\ref{tab:training}. Specifically, there are five training strategies tested, which are 
\begin{itemize}
    \item Two-stage training I (\textit{T\_1}): $\max_{\Theta'}L(\pmb{E}_t) \rightarrow \max_{\Theta'}L(\pmb{E})$
    \item Two-stage training II (\textit{T\_2}): $\max_{\Theta'}L(\pmb{E}_t)$ $\rightarrow$ $\max_{\Lambda}L(\pmb{E})$ 
    \item Two-stage training III (\textit{T\_3}): $\max_{\Theta'}L(\pmb{E}_t) \rightarrow$ $\max_{\Theta', \Lambda}L(\pmb{E})$
    \item Joint one-stage training (\textit{T\_4}): $\max_{\Theta', \Lambda}L(\pmb{E})$
    \item Three-stage training (\textit{T\_5}): $\max_{\Theta'}L(\pmb{E}_t)$ $\rightarrow$ $\max_{\Theta'}L(\pmb{E})$ $\rightarrow$ $\max_{\Lambda}L(\pmb{E})$
\end{itemize}
From the experimental results, we have the following observations. Overall, \textit{T\_2} and \textit{T\_3} achieves the best performance among all settings. Such experimental results highlight that these two are the most effective training strategies for recommendation tasks, with \textit{T\_2} being the preferred choice for balanced performance and simplicity. For high-difficulty or specialized tasks, \textit{T\_3} performs better. The choice between \textit{T\_2} and \textit{T\_3} is heavily influenced by dataset characteristics. Sparse datasets like IQON benefit from simpler strategies with fewer trainable parameters (\textit{T\_2}), while richer datasets like Polyvore can leverage the flexibility of LoRA fine-tuning (\textit{T\_3}) to achieve superior performance in challenging scenarios. For the rest of the three settings, while \textit{T\_5} applies progressive optimization, which might be effective in some other work, it falls slightly short of \textit{T\_2} and \textit{T\_3} in this study. Among all settings, \textit{T\_1} and \textit{T\_4} perform the worst, showing the importance of our first-stage pre-training on the LLM and the optimization on the projection layers for the non-textual features. 

The poor performance of joint optimization ($T_4$) stems from a misalignment between the non-textual feature projection module and the LoRA-tuned LLM. In a joint training setup, the projection module must simultaneously learn to map high-dimensional collaborative and content information into the LLM's semantic space, while the LoRA module attempts to adapt the model to specific recommendation tasks. Because these two modules start from disparate feature spaces, joint optimization leads to negative interference: the LLM essentially tries to learn task-specific reasoning based on unstable features that are in modification from an unanchored projection layer.
This interference is particularly pronounced on the Polyvore dataset due to its specific data characteristics. Polyvore outfits are typically composed of a small number of items, which—under our training sample generation strategy—results in significantly fewer Personalized Fill-In-The-Blank (P-FITB) samples compared to other datasets. In this low-sample regime, the model lacks the sufficient gradient density required to resolve the feature-space gap through joint learning, making the two-stage strategy ($T_2$ and $T_3$) essential for establishing a stable feature foundation before performing high-level task adaptation.

\begin{figure*}[th]
  \centering
  \includegraphics[width=0.65\linewidth]{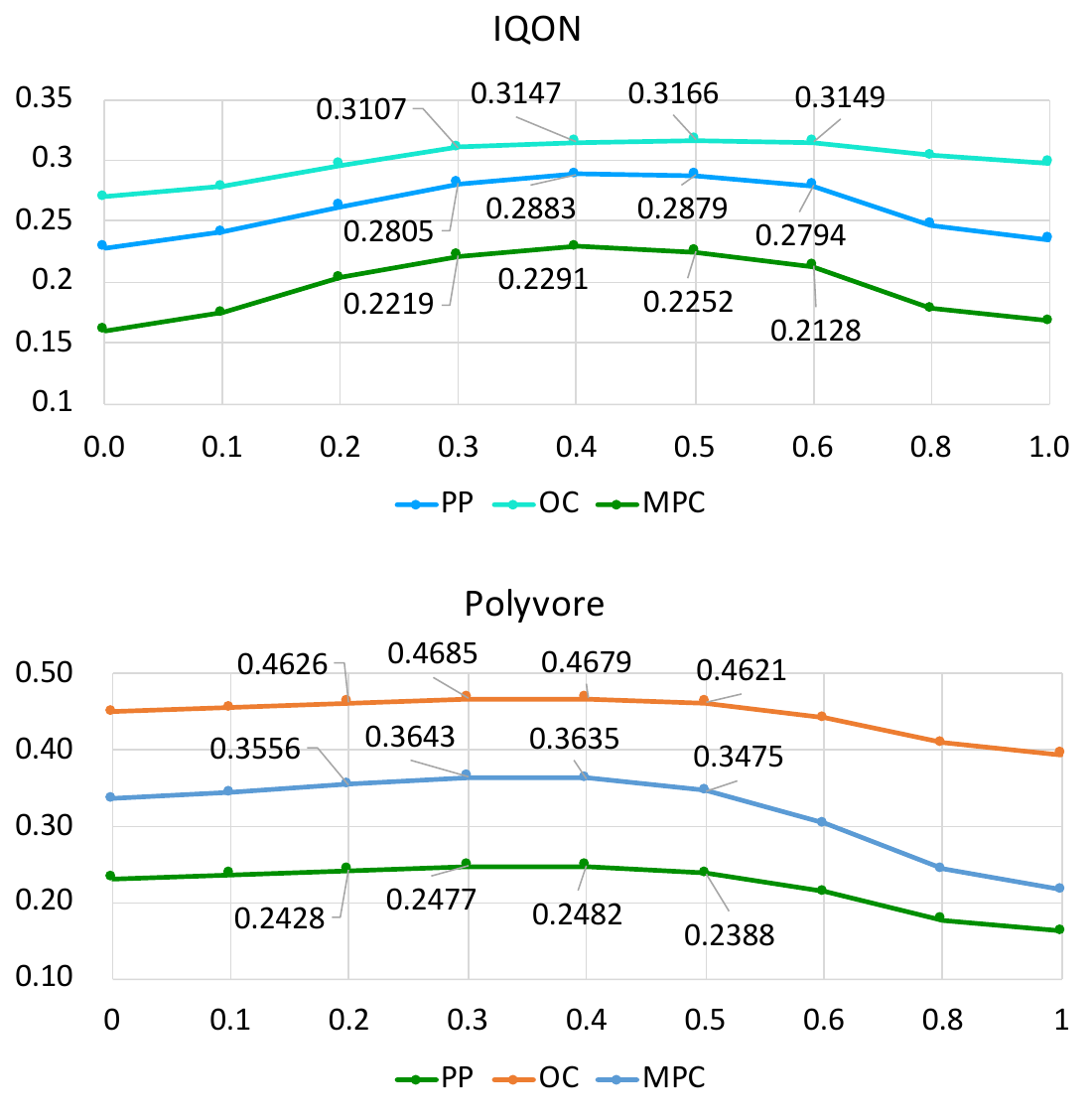}
  \caption{Personalized Outfit Generation Performance of CFALR model with different $\lambda$ settings.}
  \label{fig:cf_augmentation}
\end{figure*}

\subsection{On CF augmentation for outfit generation}
\label{sec:5.8}
In this section, we evaluate the effectiveness of outfit generation inference in our CFALR model, enhanced with CF augmentation. Using two datasets, IQON and Polyvore, we present a detailed analysis of the outfit generation performance with different levels of CF augmentation. As illustrated in Fig.~\ref{fig:cf_augmentation}. The experimental results demonstrate that incorporating CF augmentation ($\lambda \in[0,1]$) consistently outperforms the use of the fine-tuned language model alone ($\lambda=1$) across both datasets. This aligns with our earlier analysis, which suggests that while LLMs excel at handling tasks with sufficient or complex input data, they tend to struggle with tasks involving short or simple inputs, which they may infer on limited information and cannot fully explore their generalization abilities. In contrast, CF-based methods effectively directly capture interaction patterns between users and items, making them particularly valuable in scenarios where a single user or item is provided, and multiple choices must be inferred from limited information. The experimental results further reveal that optimal performance on both datasets is achieved near $\lambda =0.4$. This finding suggests that a balanced integration of the two models—LLM and CF—can effectively complement each other, leading to superior performance in our task. 
\begin{figure}[th]
  \centering
  \includegraphics[width=1\linewidth]{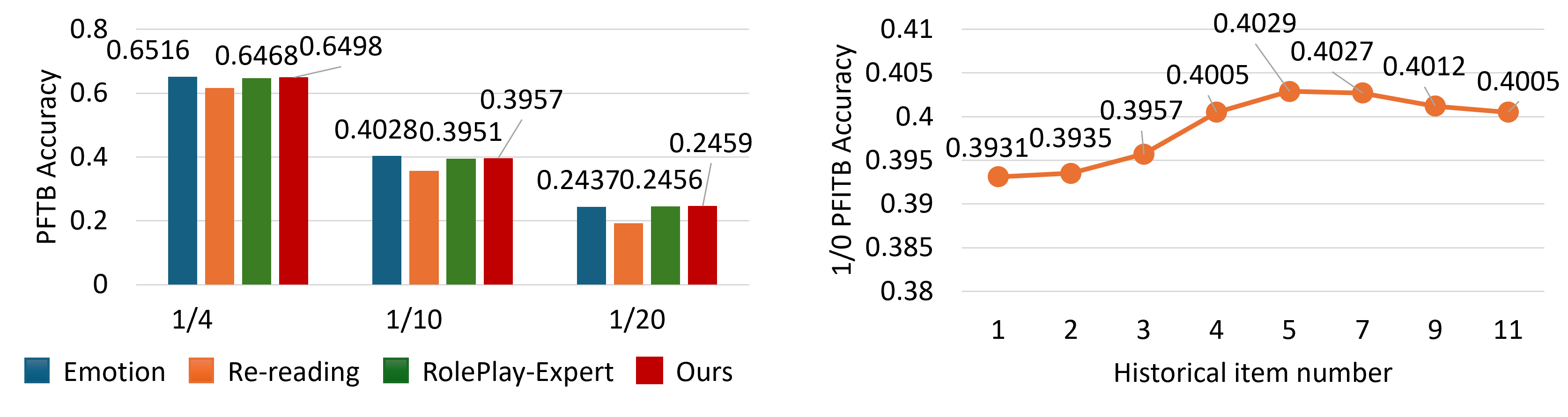}
  \caption{Personalized FITB Accuracy of CFALR with regard to different prompt template (Left) and historical item number in user encoding (Right). }
  \label{fig:prompt}
\end{figure}

\subsection{Recommendation Prompt Template and User Encoding}
We further discuss the fine-grained configurations of CFALR, specifically focusing on the prompt template for the LLM-based backbone and the number of historical items used for user encoding. Following prior studies, we evaluate four distinct prompt templates: \textit{Emotion}~\cite{li2023large}, \textit{Re-reading}~\cite{xu2024re}, \textit{RolePlay-Expert}~\cite{wang2024rolellm} and our proposed template (detailed in Sec.4.1.1). These compared templates are selected due to their demonstrated competitive performance in fashion-related recommendation tasks and their compatibility with our framework~\cite{kusano2025revisiting}. As illustrated by the experimental results on P-FITB (Fig.~\ref{fig:prompt}), the performance variance across different templates is marginal, highlighting the robustness of CFALR regarding prompt engineering. Notably, while \textit{Emotion}, \textit{RolePlay-Expert} and our template yield comparable results, we adopt our proposed template as the default setting for CFALR due to its superior computational efficiency and conciseness.  

In addition to the prompt template, we investigate the influence of the number of historical interacted items on user encoding, as illustrated in the right panel of Fig.~\ref{fig:prompt}. Our experimental results on the P-FITB task indicate a general upward trend in performance as the number of historical items increases, reaching a peak at five items (accuracy = 0.4029). However, we observe diminishing returns as the history length extends further, for instance, increasing the item number to seven, nine and eleven results in a slight performance plateau or even a marginal decrease. While the absolute difference between using one item (0.3931) and five items (0.4029) is relatively subtle, the increased history length  expands the input token count, thereby escalating the computational overhead and inference cost. Nevertheless, historical context remains indispensable to the CFALR framework. As demonstrated by the ablation study in Table~\ref{tab:feat}, removing historical items entirely causes the performance to plummet to 0.3701, which is a significant drop compared to the 0.3931 achieved with even a single historical item. This gap underscores the efficacy of historical data in helping the model identify nuanced personal preferences. To strike an optimal balance between recommendation accuracy and computational efficiency, we ultimately select five historical items as the default configuration for Polyvore and three for IQON.

\begin{table}
  \caption{Average inference time per instance for four compared methods.}
  \label{tab:inference_cost}
  \begin{tabular}{ccccc}
    \toprule
    Method &CPTM	&Vicuna-7B	&Qwen3-VL-8B	&CFALR\\
    \cmidrule(r){1-5}
    Inference time (s) &0.02	&0.12	&0.17	&0.21\\
  \bottomrule
\end{tabular}
\end{table}
 \subsection{Efficiency Analysis}
To evaluate the practical feasibility of our proposed framework, we conducted a comparative analysis of the inference latency for a single sample on the P-FITB task. As summarized in Table~\ref{tab:inference_cost}, traditional collaborative filtering (CF) methods such as CPTM exhibit the lowest latency (0.02s), benefiting from their lightweight architecture. In contrast, LLM-based models incur higher computational costs due to their massive parameter scales. Specifically, Vicuna-7B and Qwen3-VL-8B report inference times of 0.12s and 0.17s, respectively. Our proposed CFALR achieves an inference latency of 0.21s, which is slightly higher than standalone LLM backbones. This marginal increase is attributed to the multi-stage nature of our framework.

Despite the increased overhead relative to traditional CF methods, we contend that a latency of 0.21s (approximately 200ms) remains within the acceptable threshold for near real-time recommendation systems. More importantly, this trade-off is justified by the significant gains in recommendation accuracy, particularly the ability of LLMs to tackle cold-start problem that traditional, efficient models often fail to address. Moreover, with the ongoing optimization of base models and the continuous improvement of computational resources, the additional latency introduced by CFALR will become even less significant in future large-scale applications.

\subsection{Outfit Generation Analysis}
This section evaluates the performance of different outfit generation methods under varying conditions. We analyze how different outfit features—including outfit length, template, and input item category—impact the performance of our method, CFALR, compared to baselines like GPT, CPTM, and Vicuna.

\begin{figure*}
    \centering
    \includegraphics[width=1\linewidth]{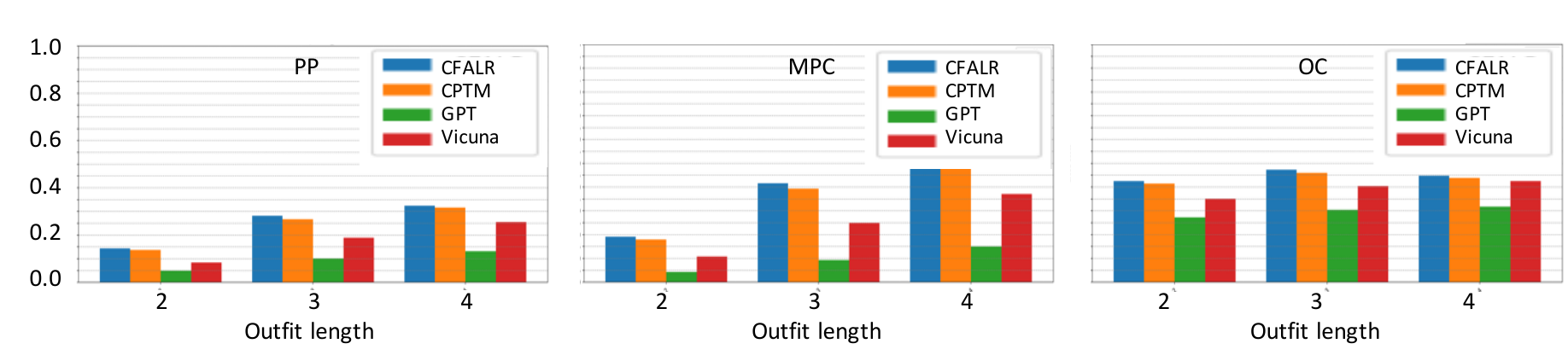}
    \caption{Outfit generation performance of four compared methods (our CFALR, GPT, CPTM and Vicuna) with regard to different length of outfit.}
    \label{fig:out_len}
\end{figure*}

Our analysis of outfit length shown in Figure~\ref{fig:out_len} reveals that PP and MPC are more sensitive to outfit length than OC. In our test dataset, outfits with fewer items are generally more challenging, resulting in lower PP and MPC scores. Consistently, our CFALR method outperforms all other methods, while GPT shows the weakest performance.

\begin{figure*}
    \centering
    \includegraphics[width=0.95\linewidth]{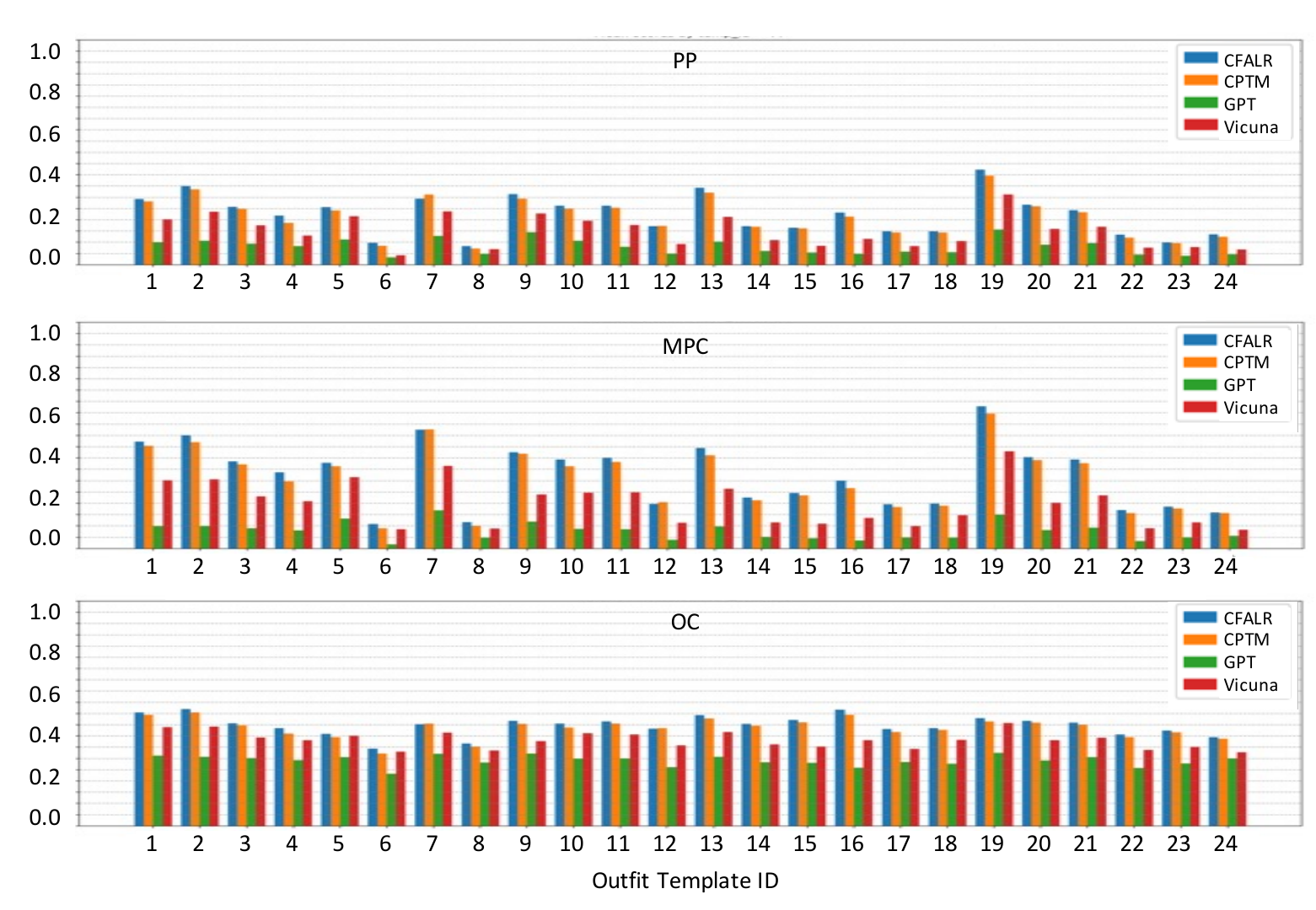}
    \caption{Outfit generation performance of four compared methods (our CFALR, GPT, CPTM and Vicuna) with regard to different template of outfit.}
    \label{fig:out_template}
\end{figure*}

\begin{figure*}
    \centering
    \includegraphics[width=0.6\linewidth]{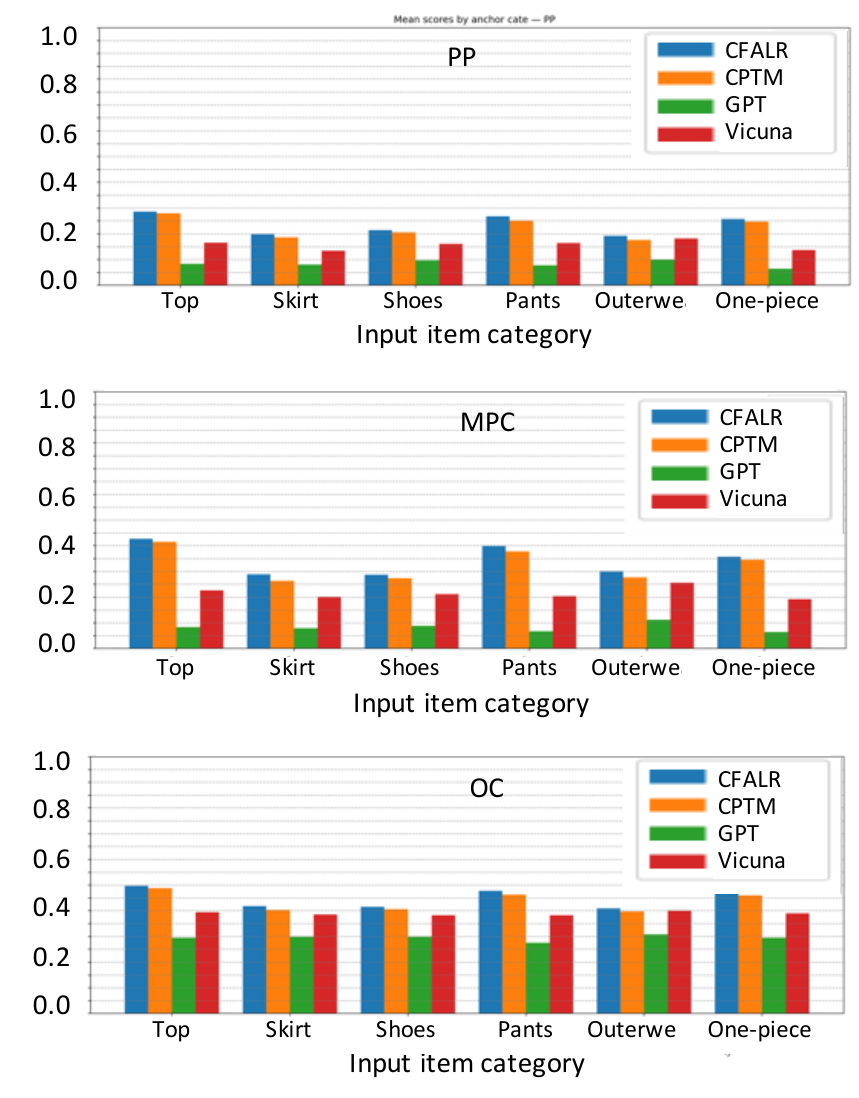}
    \caption{Outfit generation performance of four compared methods (our CFALR, GPT, CPTM and Vicuna) with regard to different input item categories.}
    \label{fig:out_input_cate}
\end{figure*}
Figure~\ref{fig:out_template} demonstrates the performance of the four methods across various outfit templates. The results highlight that all methods show significant performance variations across different outfit templates. Our CFALR method consistently outperforms all three other models on every single template tested. The PP and MPC metrics are more sensitive to template changes than OC. For all methods, performance is significantly better on certain templates (e.g., templates 7, 13, and 19) and notably worse on others (e.g., templates 6 and 8)\footnote{template id map are 1: top\_pants\_shoes, 2: pants\_outerwear\_shoes, 3: top\_skirt\_shoes, 4: top\_skirt\_outerwear, 5: top\_skirt\_outerwear\_shoes, 6: outerwear\_one-piece, 7: top\_top\_pants\_shoes, 8: female shirt\_female skirt, 9: top\_top\_pants, 10: top\_pants\_outerwear, 11: top\_outerwear\_shoes, 12: top\_pants, 13: female pants\_female sandals\_female top, 14: male bottom\_female shirt, 15: female pants\_female outerwear, 16: pants\_outerwear, 17: female shirt\_female other pants, 18: female shirt\_female jeans, 19: top\_pants\_outerwear\_shoes, 20: skirt\_outerwear\_shoes, 21: outerwear\_shoes\_one-piece, 22: skirt\_outerwear, 23: shoes\_one-piece, 24: top\_skirt}. It may suggest that some templates are easier to be comprehend by different models, therefore would achieve better performance. We also notice that for some templates CPTM performs very similar as or even outperforms our CFALR (template 3, 7, 9, 14, 21, 24). We observe that these cases are usually templates that not common or incomplete. CPTM relies more on data fitting, which therefore will not be less affected if the template is problematic. Instead, our CFALR leverages partly the power of LLM, especially its inherent knowledge, which would tend to generate outfits with more reasonable templates.

\begin{figure*}[t]
  \centering
  \includegraphics[width=0.8\linewidth]{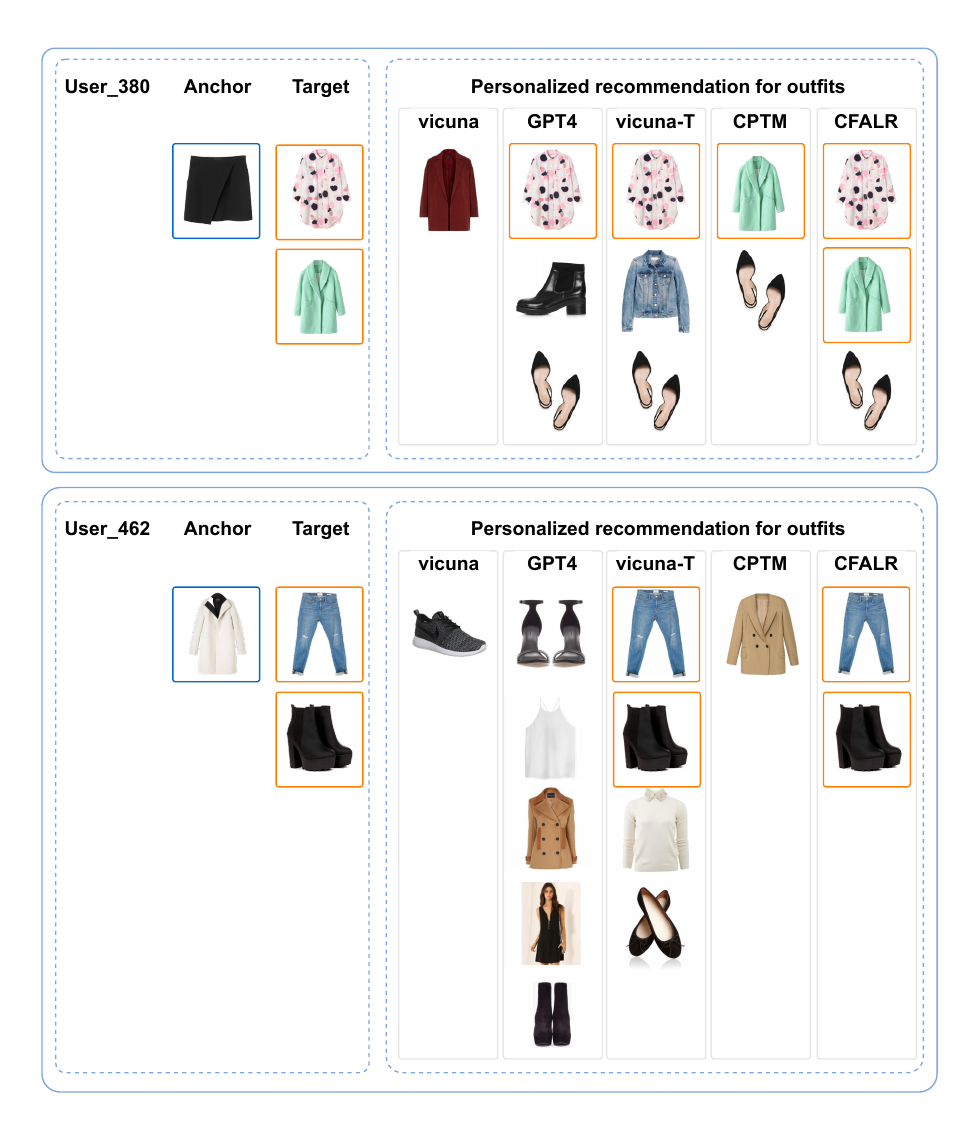}
  \caption{Two successful personalized outfit generation (POG) cases showing the results of CFALR and main compared methods.}
  \label{fig:success_case}
\end{figure*}

Figure~\ref{fig:out_input_cate} examines the impact of the input item's category on the performance of each method. The results reveal that while performance varies by input category, the differences are not as significant as those observed for outfit length or template. Overall, our CFALR and CPTM outperform the other two methods, GPT and Vicuna. All methods tend to perform better when the input item is a basic piece like a top, pants, or one-piece. In conclusion, our analysis confirms that outfit length and template are critical factors influencing the difficulty and performance of outfit generation. Our CFALR method consistently demonstrates superior performance across all tested conditions, whether judged by outfit length, template, or input item category. Its robust performance highlights its effectiveness in generating comprehensive and harmonious fashion outfits.

\subsection{Personalized Outfit Generation Cases}
Figure ~\ref{fig:success_case} provides a detailed look at two successful outfit generation (POG) cases produced by our CFALR model, alongside the outputs from major comparative methods. This visual analysis highlights CFALR’s key advantages and the distinct limitations of the competing models. In both cases, CFALR consistently demonstrates a superior ability to generate outfits that not only align with the user's preferences but also maintain a high degree of consistency and compatibility with the anchor item and other selected pieces. This strength underscores CFALR's effectiveness in producing complete and coherent outfits that go beyond simple item association.

The first case provides a clear example of the efficacy of our CF-augmentation scheme. Here, CFALR correctly predicts two items by strategically integrating probabilities from both the fine-tuned Vicuna and the CPTM model. This successful outcome demonstrates that our approach outperforms individual models by leveraging their complementary strengths. In contrast, the comparative methods reveal significant inherent limitations. For instance, the off-the-shelf Vicuna model struggles with outfit generation, producing incomplete sets containing only a single item.

Furthermore, we observe that vanilla LLMs are plagued by positional bias, an inherent issue where models are sensitive to the sequential order of input candidates~\cite{wang2024large}. While GPT-4 appears less sensitive based on our empirical observation, Vicuna exhibits a severe preference for selecting the first item from a given list, which compromises the diversity and relevance of its outputs. This is a critical challenge of LLMs for recommendation tasks. Notably, we find that after fine-tuning with the P-FITB (Personalized Fill-In-The-Blank) task, the model effectively addresses this bias. This suggests that P-FITB successfully trains the LLM to prioritize compositional relations between items over their sequential positions.

While vanilla Vicuna outputs too few items, the pure CF-based method, CPTM, also tends to generate short outfits. Because CPTM focuses primarily on item-wise and user-item compatibility, it lacks the holistic understanding required to ensure the completeness of a generated group. Conversely, while GPT-4 and the fine-tuned Vicuna-T generate longer outfits, they often fail to maintain a coherent composition. They frequently produce redundant or nonsensical items—such as two separate footwear items in a single outfit—revealing a lack of hard, rule-based fashion knowledge. Our CF-augmentation mitigates this by restoring these structural constraints, ensuring that CFALR achieves superior accuracy, diversity, and alignment with practical fashion logic. 
The CF augmentation scheme is particularly impactful, as it significantly enhances the quality and relevance of the generated outfits.

\begin{figure*}[t!]
  \centering
  \includegraphics[width=0.75\linewidth]{ 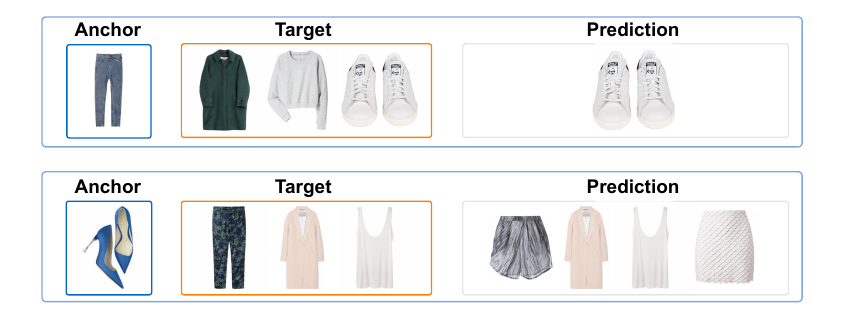}
  \caption{Two failure POG cases of CFALR.}
  \label{fig:failure_case}
\end{figure*}

Figure~\ref{fig:failure_case} provides a candid analysis of two failure cases for our CFALR model, highlighting specific limitations and areas for future improvement. In the first case, the model fails to generate a complete outfit, predicting only one additional item. While this single item aligns well with the user's preferences and correctly matches the anchor item, the output ultimately fails to fulfill the primary task of generating a full and coherent outfit. This suggests a potential weakness in the model's generation strategy, indicating a need for a more robust mechanism to ensure outfit completeness. The second case reveals a different type of failure. Here, CFALR successfully generates multiple items but includes a significant redundancy, such as two distinct pieces of bottom wear. This suggests that while the model can produce a longer sequence of items, it occasionally lacks a strategic understanding of outfit composition and violates fundamental fashion rules. This limitation points to the need for enhanced constraints to ensure better category diversity and overall balance within the generated outfits.

Despite these shortcomings, it is important to note that both failure cases still demonstrate a core strength of the CFALR model: its ability to provide appropriate, context-aware item-level suggestions that align with user preferences. The failures are not due to a complete misunderstanding of the user's intent, but rather to challenges in the final steps of outfit completion and compositional logic. Addressing these issues will be the focus of future work to improve the model's overall performance and reliability.

\section{Conclusion and Future Work}
This paper presents Collaborative Filtering-Augmented Large Language Model for Recommendation (CFALR), a novel fashion outfit recommendation method and the first LLM-based approach designed to address the limitations of traditional methods that heavily depend on dense user-item interaction data. Built upon the open-source pre-trained LLM Vicuna, CFALR was fine-tuned with outfit recommendation data using a personalized fill-in-the-blank problem formulation. The proposed method leverages the strengths of collaborative filtering to capture implicit interaction patterns by employing a hybrid encoding for users and items. Additionally, a CF augmentation scheme was introduced to integrate the predicted probabilities of the CF and the language model, achieving a balance between interaction pattern modeling and semantic understanding. Experimental results on two benchmark fashion outfit recommendation datasets demonstrate that CFALR excels in delivering personalized and compatible item-level recommendations while also generating cohesive and user-specific outfits. 

In future work, we aim to focus on several directions to further enhance this study. First, we will enhance the user modeling, trying to develop a more sophisticated way to better capture user's fashion tastes or intent. Second, instead of simply aligning the visual and collaborative filtering features in the language model backbone through target objective-specific fine-tuning , we will explore better integration method with better understanding of the domain specific data and knowledge. Third, with this model as a foundation, we will develop an interactive outfit recommendation framework that allows user feedback and revision on the recommendation results.

\bibliographystyle{ACM-Reference-Format}
\bibliography{refe}

\end{document}